\documentclass[aps,prd,reprint,superscriptaddress,preprintnumbers,longbibliography]{revtex4-2}
\usepackage{amsmath,amssymb,bm,mathrsfs}
\usepackage[sort&compress]{natbib}
\usepackage{srcltx}
\usepackage{dsfont}
\usepackage{epsfig}
\usepackage{slashed}
\usepackage{bbold}
\usepackage{psfrag}
\usepackage[dvipsnames]{xcolor}
\PassOptionsToPackage{caption=false}{subfig}
\usepackage{subfig}
\usepackage{xfrac}
\usepackage{multirow}
\usepackage{booktabs}
\usepackage[colorlinks=true
,urlcolor=blue
,anchorcolor=blue
,citecolor=blue
,filecolor=blue
,linkcolor=red
,menucolor=blue
,linktocpage=true
,pdfproducer=medialab
,pdfa=true
]{hyperref}
\usepackage{cleveref}
\usepackage{setspace}
\usepackage{relsize}
\usepackage{cancel}
\usepackage{tikz,tikz-feynman,enumerate,comment}
\usepackage{amsmath,mathtools}
\usepackage{comment}
\usepackage[paperwidth=210mm,paperheight=297mm,centering,hmargin=2cm,vmargin=2.48cm]{geometry}

\usepackage[normalem]{ulem}
\usepackage[caption=false]{subfig}

\newcommand{\ie}{\textsl{i.e.}~}

\newcommand{\order}[1]{\mathcal{O}\!\left(#1\right)}

\newcommand{\Tr}{\mathrm{Tr}}

\newcommand{\dd}{\mathrm{d}}

\newcommand{\ee}{e}

\newcommand{\boldmathsymbol}[1]{{\ensuremath{\boldsymbol{#1}}}}

\newcommand{\calH}{\mathcal{H}}

\newcommand{\Rea}{\Re \mathrm{e}\,}
\newcommand{\Ima}{\Im \mathrm{m}\,}

\newcommand{\Real}[1]{\Rea \left[ #1 \right] }
\newcommand{\Imag}[1]{\Ima \left[ #1 \right] }

\newcommand{\cs}{c_{_\mathrm{S}}}

\newcommand{\efolds}{$e$-folds~}

\newcommand{\beq}{\begin{equation}}
	\newcommand{\eeq}{\end{equation}}
\newcommand{\bea}{\begin{equation}\begin{aligned}}
		\newcommand{\eea}{\end{aligned}\end{equation}}

\newlength{\wsingfig}
\newlength{\wdblefig}
\newlength{\wquadfig}
\newlength{\wtriplefig}
\newcommand{\Eq}[1]{Eq.~(\ref{#1})}
\newcommand{\Eqs}[1]{Eqs.~(\ref{#1})}
\newcommand{\Fig}[1]{Fig.~{\ref{#1}}}

\newcommand{\Refa}[1]{Ref.~{\cite{#1}}}
\newcommand{\Refs}[1]{Refs.~{\cite{#1}}}
\newcommand{\Sec}[1]{Sec.~\ref{#1}}

\newcommand{\bs}[1]{\boldsymbol{#1}}

\begin{document}


\title{Quantum recoherence in the early universe}

\author{Thomas Colas}
\affiliation{Universit\'e Paris-Saclay, CNRS, Institut d'Astrophysique Spatiale, 91405, Orsay, France}
\affiliation{Laboratoire Astroparticule et Cosmologie, Universit\'e Denis Diderot Paris 7, 10 rue Alice Domon et L\'eonie Duquet, 75013 Paris, France}

\author{Julien Grain}
\affiliation{Universit\'e Paris-Saclay, CNRS, Institut d'Astrophysique Spatiale, 91405, Orsay, France}

\author{Vincent Vennin}
\affiliation{Laboratoire de Physique de l'Ecole Normale Sup\'erieure, ENS, CNRS, Universit\'e PSL, Sorbonne Universit\'e, Universit\'e Paris Cit\'e, F-75005 Paris, France}
\affiliation{Laboratoire Astroparticule et Cosmologie, Universit\'e Denis Diderot Paris 7, 10 rue Alice Domon et L\'eonie Duquet, 75013 Paris, France}

\begin{abstract}
\noindent Despite being created through a fundamentally quantum-mechanical process, cosmological structures have not yet revealed any sign of genuine quantum correlations. 
Among the obstructions to the direct detection of quantum signatures in cosmology, environmental-induced decoherence is arguably one of the most inevitable. 
Yet, we discover a mechanism of quantum recoherence for the adiabatic perturbations when they couple to an entropic sector. 
After a transient phase of decoherence, a turning point is reached, recoherence proceeds and adiabatic perturbations exhibit a large amount of self-coherence at late-time.
This result is also understood by means of a non-Markovian master equation, which reduces to Wilsonian effective-field theory in the unitary limit. 
This allows us to critically assess the validity of open-quantum-system methods in cosmology and to highlight that re(de)coherence from linear interactions has no flat-space analogue.
\end{abstract}

\maketitle

%
Our current understanding of cosmology traces back the origin of structures to quantum fluctuations in the primordial vacuum.  
Not only inflation, the leading scenario \cite{Starobinsky:1979ty,Guth:1980zm, Starobinsky:1980te,Sato:1980yn, Linde:1981mu, Mukhanov:1981xt, Guth:1982ec, Albrecht:1982wi, Starobinsky:1982ee, Hawking:1982cz, Linde:1983gd, Bardeen:1983qw, Mukhanov:1988jd}, but also most alternatives \cite{Khoury:2001wf, Brandenberger:2016vhg, 2016arXiv161201236A, Barrau:2013ula} rely on this mechanism.
However, whether or not one can prove (or disprove) the quantum origin of cosmological inhomogeneities remains an open issue~\cite{Campo:2005sv, Martin:2015qta, Maldacena:2015bha, Martin:2016tbd, Martin:2016nrr, Kanno:2016gas, Choudhury:2016cso, Martin:2017zxs, Green:2020whw, Ando:2020kdz, Espinosa-Portales:2022yok, Micheli:2022tld}. 
Independently of the observational challenge it may constitute, it is generally argued \cite{Brandenberger:1990bx, Brandenberger:1992sr, Barvinsky:1998cq, Lombardo:2005iz, Kiefer:2006je, Martineau:2006ki, Burgess:2006jn, Prokopec:2006fc, Kiefer:2008ku, Nelson:2016kjm, Hollowood:2017bil, Martin:2018zbe, Martin:2018lin, Kanno:2020usf, Martin:2021znx, DaddiHammou:2022itk, Burgess:2022nwu} that any genuine quantum signature is likely to be erased by the quantum decoherence~\cite{Zurek:1981xq, Zurek:1982ii, Joos:1984uk} induced by environmental degrees of freedom.
This is why studying decoherence channels \cite{Koks:1996ga, 
Campo:2008ju, Campo:2008ij, Anastopoulos:2013zya, Fukuma:2013uxa, Akhtar:2019qdn, Kaplanek:2020iay, 2020JHEP...03..008K, Brahma:2021mng, Burgess:2021luo, Oppenheim:2022xjr, Brahma:2022yxu} has become of primary importance to assess the severity of this potential obstruction.
Recent progresses in the cosmological open-quantum-system program provide this line of investigation with a robust toolbox, which nonetheless needs to be adapted and benchmarked since cosmology tends to break some of the assumptions it otherwise rests on~\cite{Shandera:2017qkg, Kaplanek:2022xrr, Colas:2022hlq}.
In this Letter, we investigate the decoherence process in arguably one of the most generic extensions to single-field slow-roll inflation \cite{Achucarro:2010da, Cespedes:2012hu, Achucarro:2012sm, Assassi:2013gxa, Tong:2017iat, Pinol:2020kvw, Pinol:2021aun, Jazayeri:2022kjy}.
Contrary to common wisdom, we discover that, after a transient phase of decoherence, \emph{recoherence} takes place and the final state exhibits large levels of self-coherence. 
Notably, this result has no flat-space analogue. 
Heavy fields are ubiquitous when inflation is embedded in high-energy constructions, both from a model-building perspective~\cite{Turok:1987pg, Damour:1995pd, Kachru:2003sx, Krause:2007jk, Chen:2009zp, Tolley:2009fg, Chen:2012ge, Pi:2012gf, Baumann:2014nda, Pinol:2021aun, Pinol:2020kvw, Jazayeri:2022kjy} and from an effective-field-theory (EFT) approach~\cite{Achucarro:2010da, Shiu:2011qw, Cespedes:2012hu, Achucarro:2012sm, Assassi:2013gxa, Tong:2017iat}.
From a bottom-up viewpoint, the dynamics of the fluctuations in the adiabatic direction $\zeta$ and the entropic direction $\mathcal{F}$ is given at linear order by~\cite{Assassi:2013gxa}
\begin{align}\label{eq:eq1}
\mathcal{L} =&  a^2  \epsilon M_{\mathrm{Pl}}^2 \zeta^{\prime2} - a^2 \epsilon M_{\mathrm{Pl}}^2  \left(\partial_i \zeta\right)^2 \nonumber 
+ \frac{1}{2}a^2 \mathcal{F}^{\prime2} \\
& - \frac{1}{2}a^2 \left(\partial_i  \mathcal{F}\right)^2
- \frac{1}{2}m^2 a^4 \mathcal{F}^2 - \rho a^3 \sqrt{2\epsilon} 
 M_{\mathrm{Pl}} \zeta^{\prime} \mathcal{F}.
\end{align}
The Lagrangian density $\mathcal{L}$ is expressed in conformal time $\eta$, primes denote derivative with respect to $\eta$ and $\left(\partial_i \zeta\right)^2 \equiv \delta_{ij}\partial_i \zeta \partial_j \zeta$.
Finally, $a$ is the scale factor, $\epsilon$ the first slow-roll parameter and $M_{\mathrm{Pl}}$ the Planck mass.
The massless degree of freedom $\zeta$ is the curvature perturbation and is directly observed in the cosmic microwave background (CMB)~\cite{Aghanim:2018eyx,Planck:2018jri} and the large-scale structure of the universe~\cite{SDSS:2005xqv, BOSS:2014hwf,Colas:2019ret,DES:2022qpf}.
The coupling $\rho$ corresponds to the rate of turn in field space and mixes the adiabatic and entropic directions.
It is constant at leading order in slow roll~\cite{Chen:2009zp,Tolley:2009fg, Chen:2012ge,Pi:2012gf}, and when a specific model is considered, it can be related to its microphysical parameters~\cite{Achucarro:2010da,  Cespedes:2012hu, Achucarro:2012sm}.
From an EFT perspective, $ \zeta^{\prime} \mathcal{F}$ is the only operator compatible with the shift symmetry of the Goldstone mode and with spatial homogeneity of the background~\cite{Assassi:2013gxa}, hence \Eq{eq:eq1} captures the leading effect in the derivative expansion of generic multiple-field models~\cite{Pinol:2021aun}.

This setting has mostly been studied from a phenomenological perspective, \ie focusing on calculations of the observable power spectrum using the in-in formalism~\cite{Chen:2009zp,Tolley:2009fg,Chen:2012ge,Pi:2012gf} or by means of single-field Wilsonian EFTs~\cite{Achucarro:2010da,Cespedes:2012hu, Achucarro:2012sm, Assassi:2013gxa, Tong:2017iat}. The latter approaches incorporate unitary effects only, hence they cannot describe decoherence~\cite{Breuer:2002pc}.
In this Letter, we make this possible by treating \Eq{eq:eq1} within the open-quantum-system framework and by extracting quantum-information-theoretic properties of the curvature perturbations.
This leads us to the phenomenon of quantum recoherence.
We solve the problem both exactly and using an effective approach, allowing us to better assess the performance of such methods in a cosmological context.
%

\begin{figure*}[!t]
\centering
\includegraphics[width=1\textwidth]{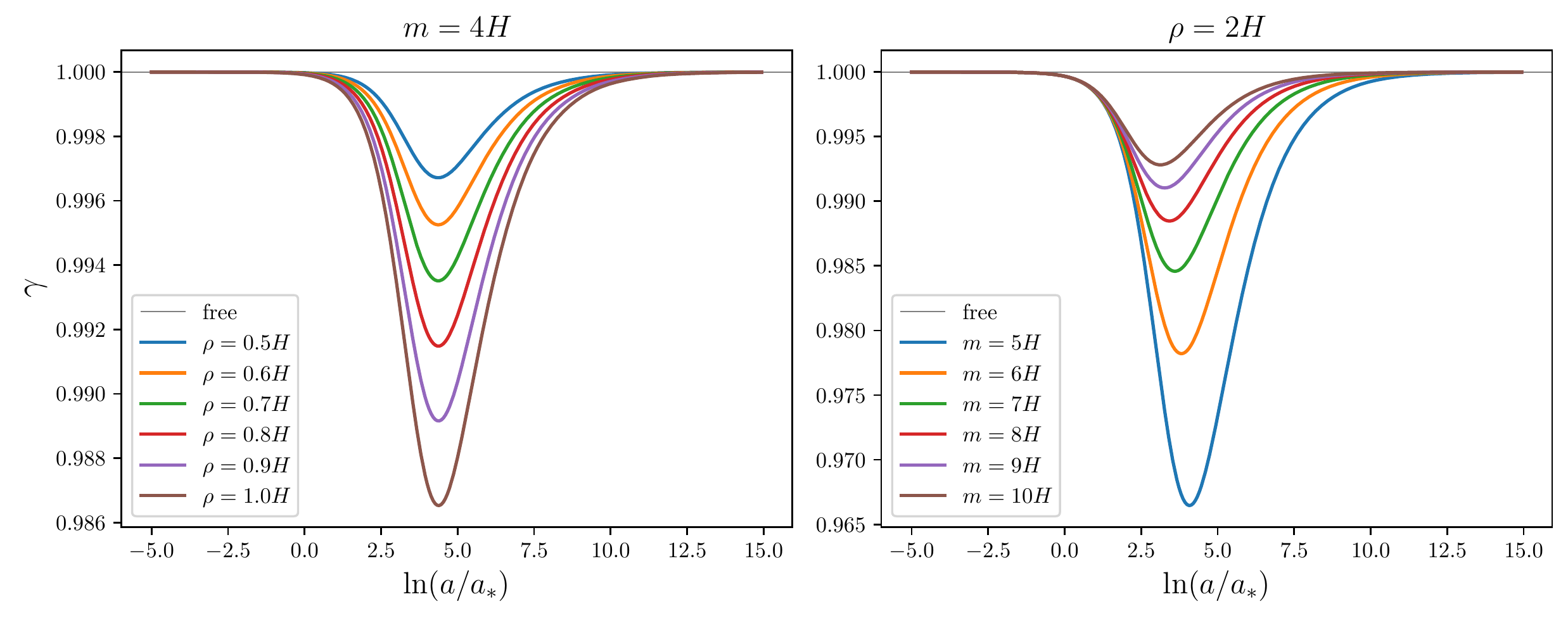}
\vspace{-0.8cm}
\caption{State purity $\gamma$ as a function of the number of efolds $\ln(a/a_*)$ since Hubble exit of the scale $k=a_* H$, for a few values of the coupling parameter $\rho$ (left panel) and of the entropic mass $m$ (right panel). After a transient phase of decoherence,  the purity reaches a minimum and increases again. This recoherence phenomenon yields high levels of self-coherence at late-time. In practice, Eqs.~(S10) and (S11) of the SM are integrated with Bunch-Davies initial conditions from $\ln(a/a_*) = -15$ to $\ln(a/a_*) = 15$ and for constant $H$. 
}
\vspace{-0.5cm}
\label{fig:reco}
\end{figure*}

\vspace{0.2cm}
\noindent {\bf Quantum recoherence.}
%
From an open-quantum-system perspective, our goal is to describe the dynamics of the adiabatic sector (the \textit{system}) once the heavy entropic direction (the \textit{environment}) has been integrated out.
\Eq{eq:eq1} being quadratic, different Fourier modes decouple on a homogeneous background, which allows us to focus on a given mode $\bs{k}$.  
In the asymptotic past, this mode is placed in the Bunch-Davies vacuum state~\cite{Bunch:1978yq}, described by the Gaussian density matrix $\widehat{\rho}_{0}$, and the linearity of the dynamics preserves the Gaussianity of the state \cite{PhysRevA.36.3868, SIMON1987223, Colas:2021llj}.
Hence, the state of the system is entirely characterised by the covariance $\boldmathsymbol{\Sigma}_{ij}(\eta) \equiv \frac{1}{2} \Tr \left\{ [ \boldmathsymbol{\widehat{z}}_{\zeta,i}(\eta)  \boldmathsymbol{\widehat{z}}_{\zeta,j}(\eta) + \boldmathsymbol{\widehat{z}}_{\zeta,j}(\eta)  \boldmathsymbol{\widehat{z}}_{\zeta,i}(\eta) ] \widehat{\rho}_{0}\right\}$ where $\widehat{\bs{z}}_{\zeta} \equiv (\widehat{v}_{\zeta}, \widehat{p}_{\zeta})^{\mathrm{T}}$ contains the configuration and momentum operators of the Mukhanov-Sasaki variable $v_{\zeta} \equiv - a \sqrt{2\epsilon} M_{\mathrm{Pl}} \zeta$.
In the Supplemental Material (SM), from \Eq{eq:eq1} we derive an exact equation of motion for $\boldmathsymbol{\Sigma}_{ij}(\eta)$, known as a transport equation~\cite{ehrenfest_bemerkung_1927,Werth:2023pfl, Raveendran:2023dst}. 
When integrated numerically, the power spectra one obtains are well reproduced by standard EFT results~\cite{Chen:2012ge,Pi:2012gf,Cespedes:2012hu} in the regime $m \gg H$, as is shown in the SM. 
In particular, we recover that the main effect from the heavy field is a simple rescaling, proportional to $\rho^2/m^2$, of the amplitude of the scale-invariant power spectrum of $\zeta$.
This rescaling is however degenerate with other single-field effects such as a reduced speed of sound~\cite{Chen:2009zp,Pajer:2013ana,Kenton:2015lxa}, so it cannot be used to reveal the existence of an environment.

The state being Gaussian, the covariance $\boldmathsymbol{\Sigma}(\eta)$ not only contains all observables of the adiabatic sector but also fully specifies its quantum state, \ie the reduced density matrix $\widehat{\rho}_{\mathrm{red}} \equiv \Tr_{\mathcal{F}} (\widehat{\rho})$ where the entropic degrees of freedom are traced over.
This allows us to study quantum properties of  $\widehat{\rho}_{\mathrm{red}}$, in particular the transition from a pure quantum state into a statistical mixture due to the interaction with an environment \cite{Zurek:1981xq, Zurek:1982ii, Joos:1984uk}.
This transition is assessed by the so-called purity parameter~\cite{Serafini:2003ke, Walschaers:2021zvx} $\gamma \equiv \Tr\left(\widehat{\rho}_{\mathrm{red}}^2\right)$, which equals one if the state is pure and is smaller than one otherwise.
The system is said to have decohered when $\gamma\ll 1$, with $\gamma=0$ corresponding to a maximally mixed state. 
The link between the numerical value of $\gamma$ and the erasure of explicit quantum signatures (such as Bell inequality violations) has been investigated in \Refa{Martin:2022kph} for the class of states considered in this work.
Note that $\gamma$ remains invariant under reparametrisation of the canonical variables~\footnote{Single-mode Gaussian systems are fully characterised by a unique symplectic invariant (\ie independent of the choice of canonical variables used to parametrise phase space). Any (symplectic-invariant) decoherence tracer, such as the entanglement entropy, thus follows the same trend as purity.}, and for a Gaussian state one simply has $\gamma(\eta)  = \frac{1}{4} \det\left[\boldmathsymbol{\Sigma}(\eta)\right]^{-1} $~\cite{2003PhRvA..68a2314P, 2014arXiv1401.4679A}. 

\begin{figure}[!t]
\centering
\includegraphics[width=0.49\textwidth]{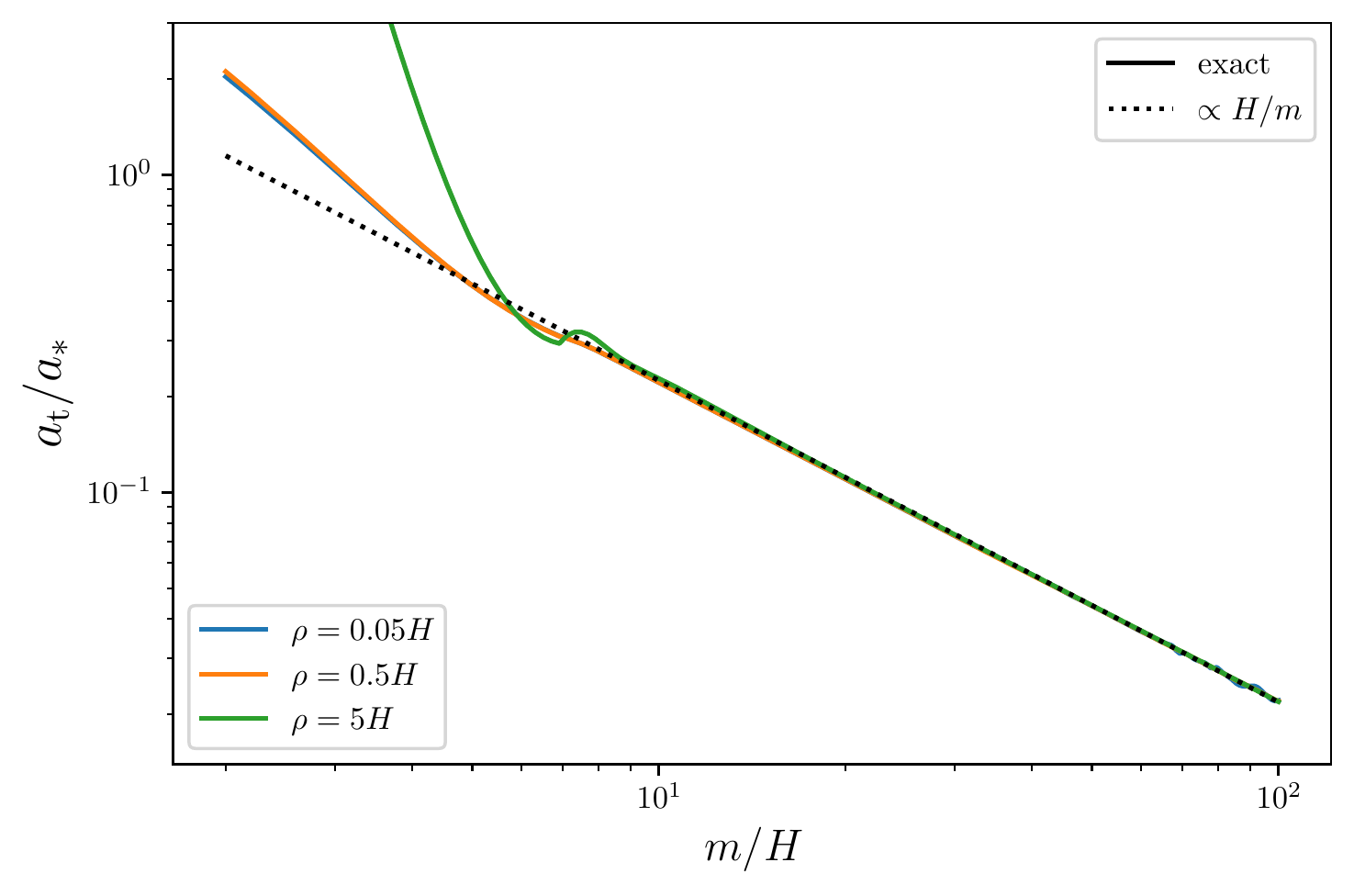}
\vspace{-0.8cm}
\caption{
Turning point for the purity: value of the scale factor $a_{\mathrm{t}}$ at which the purity starts increasing, as a function of $m$ and for a few values of $\rho$. When $m\gg H$, $a_{\mathrm{t}} \simeq 2.3 k/m$, which corresponds to the Compton wavelength of the entropic field.
}
\vspace{-0.4cm}
\label{fig:fig1bis}
\end{figure}

In \Fig{fig:reco}, we display the purity parameter $\gamma$ as a function of the number of efolds $\ln(a/a_*)$ since Hubble exit of the scale $k=a_* H$ under consideration, where at leading order in slow roll $H$ is constant.
After a transient phase of decoherence, a turning point occurs and recoherence (\ie growing $\gamma$) takes place, with large levels of self-coherence at late time. 
For heavy masses $m \gg H$, the turning point occurs in the sub-Hubble regime, when the scale $k$ crosses the Compton wavelength of the entropic field $1/m$, as shown in \Fig{fig:fig1bis}. 
The departure from a pure state increases with $\rho$ and decreases with $m$, in agreement with the EFT intuition that heavier environments leave a smaller imprint on light degrees of freedom. 
At late time, one can expand the transport equations in the super-Hubble limit and in the SM we find that, at leading order in $\rho^2$,
\begin{align}\label{eq:eq4}
\gamma &= \gamma_{\infty} -  \frac{\rho^2}{m^2}  \frac{k}{am}\, .
\end{align}
This confirms that the purity does increase at late time for all super-Hubble scales, at a rate controlled by the ratio between the Compton wavelength and the mode wavelength. Thus it quickly reaches the asymptotic value $\gamma_{\infty} < 1$, since for the scales probed in the CMB, $k/(am)$ is typically of order $\ee^{-50} H/m$.
%

The occurrence of recoherence might seem surprising in the light of previous works on decoherence in this model~\cite{Prokopec:2006fc, Raveendran:2022dtb} and in other cosmological scenarios~\cite{Koks:1996ga, Prokopec:2006fc, Anastopoulos:2013zya, Fukuma:2013uxa, Akhtar:2019qdn, Kaplanek:2020iay, 2020JHEP...03..008K, Brahma:2021mng, Burgess:2021luo, Oppenheim:2022xjr, Brahma:2022yxu, Colas:2022hlq}.
One may indeed expect that, once information about the system has ``leaked'' into the environment, it cannot ``come back''. Yet we argue that there is no contradiction with the existing literature.
This is due to the small effective size of the environment here: the system couples to a single Fourier mode of the environment.
This implies that, contrary to the open quantum systems usually considered, the environment does not behave as a thermal bath~\cite{Breuer:2002pc}.

To gain further insight into the finite-environment effects to be expected in this model, one may consider its analogue in Minkowski spacetime.
When the background is static, linear interactions can only induce mixing between the light and heavy sectors so that the purity exhibits oscillations at frequencies given by the characteristic timescales of the system and the environment, as checked explicitly in the SM.
If the coupling is quenched off, oscillations stop and the purity freezes at the time of the quench. 
In de-Sitter spacetime, this is precisely what happens, since the non-trivial background dynamics makes the coupling effectively time dependent.

\begin{figure}[!t]
\centering
\includegraphics[width=0.49\textwidth]{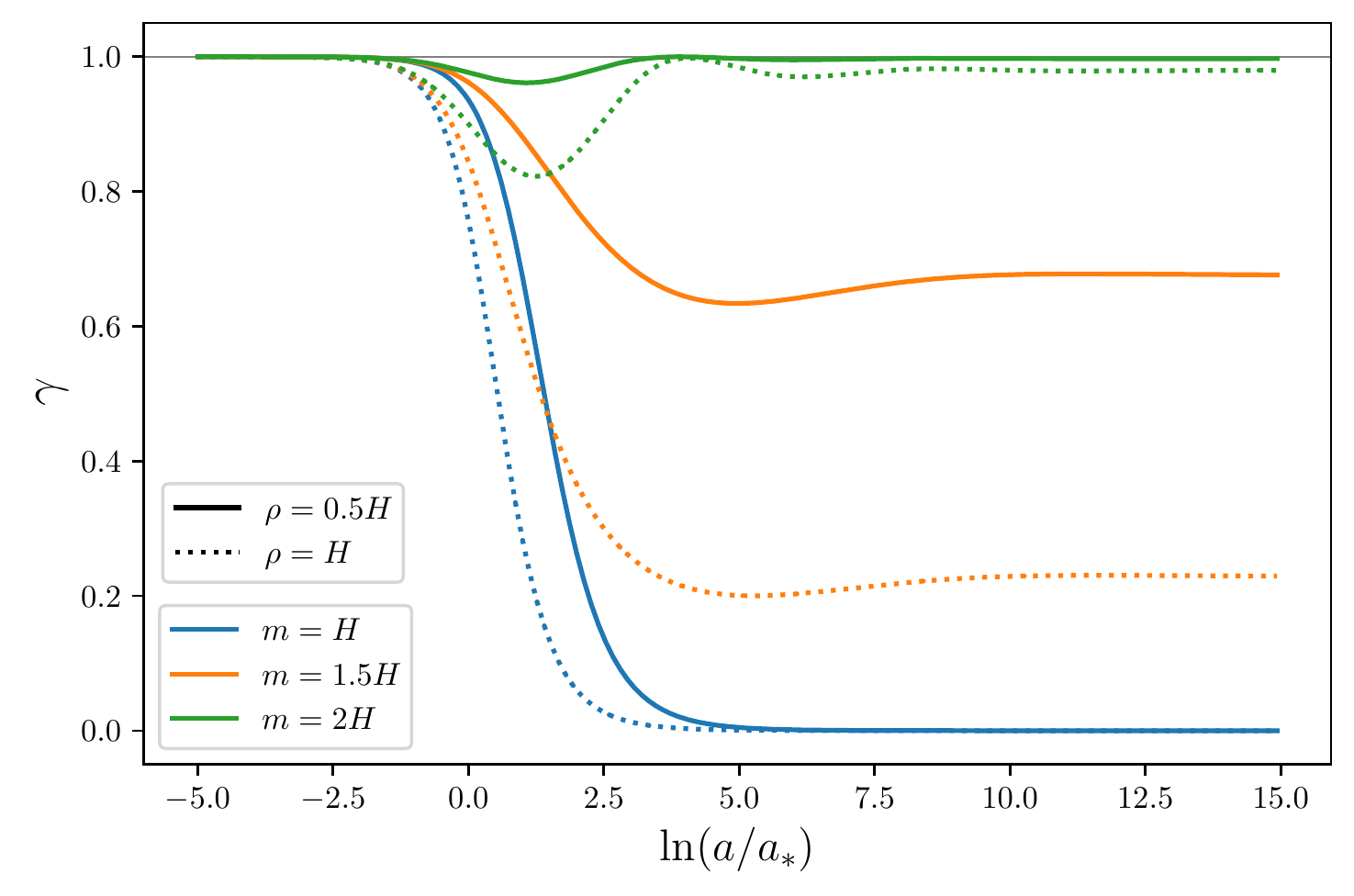}
\vspace{-0.8cm}
\caption{
Same as in \Fig{fig:reco} for lighter environments. At late time, one either observes recoherence ($m>3H/2$), purity freezing ($m\simeq 3H/2$, with an asymptotic value that strongly depends on $\rho$) or decoherence ($m<3H/2$).
}
\vspace{-0.4cm}
\label{fig:fig2}
\end{figure}

This can be seen in \Fig{fig:fig2} where small entropic masses are used to better highlight the following stages in the evolution of purity. 
When $k \gg am$, the mode functions of both fields oscillate at the same frequency $k/a$, in their vacuum state. 
Then $am \gg k \gg aH$ and the two frequencies differ: the system oscillates at frequency $k/a$ while the environment oscillates at frequency $m$, hence the purity oscillates as in flat space.
Finally, when $k \ll aH$, two behaviours can be observed, depending on $m/H$. 
If $m > \frac{3}{2}H$, entropic perturbations are heavy hence they oscillate and quickly decay~\cite{Cespedes:2012hu}. Since $\zeta' $ also decays as $1/a^2$ on super-Hubble scales~\cite{Lyth:2004gb} (this is the so-called ``decaying mode''), the coupling between adiabatic and entropic perturbations is effectively turned off. This is why the value of the purity freezes (see the cases $m=1.5 H$ and $m=2H$ in \Fig{fig:fig2}). 
When the environment is lighter, $\mathcal{F}$ acquires a growing mode that keeps the interaction term $\zeta' \mathcal{F}$ active in spite of the decay of $\zeta'$. This leads to decoherence (see the case $m=H$ in \Fig{fig:fig2}), driven by the dynamics of the expansion. This is similar to the setup studied in \Refa{Prokopec:2006fc}, where an additional $\zeta \mathcal{F}$ interaction term is considered that is not suppressed by the decaying mode $\zeta'$ on large scales, and to the cosmological two-field model investigated in \Refa{Colas:2022hlq}. 
In all these cases, the system is driven into a mixed state by the dynamical generation of entangled pairs of quanta between $\zeta$ and $\mathcal{F}$, which explains why decoherence takes place in spite of the environment being effectively made of one single degree of freedom.


In the present setting, the entropic direction is typically expected to be heavy, but it is interesting to see that, formally, by varying $m$, one interpolates between these three possible outcomes: recoherence, purity freezing and decoherence. Note that the intermediate mass $m \simeq \frac{3}{2}H$ is also of phenomenological interest in the context of quasi-single field inflation \cite{Chen:2009zp,Tolley:2009fg,Chen:2012ge,Pi:2012gf, Assassi:2013gxa}, and that the fate of the purity in that case is particularly sensitive to $\rho$, see \Fig{fig:fig2}.
%

\vspace{0.2cm}
\noindent {\bf A master-equation treatment.}
%
The model~\eqref{eq:eq1} being linear, it can be solved exactly but this is in general not possible. This is why open quantum systems are usually approached with effective methods known as master equations. We now apply such methods to the present setup, in order to check their validity, and to shed additional light on the imprint left by $\mathcal{F}$ on $\zeta$.

\begin{figure}[!t]
\centering
\includegraphics[width=0.49\textwidth]{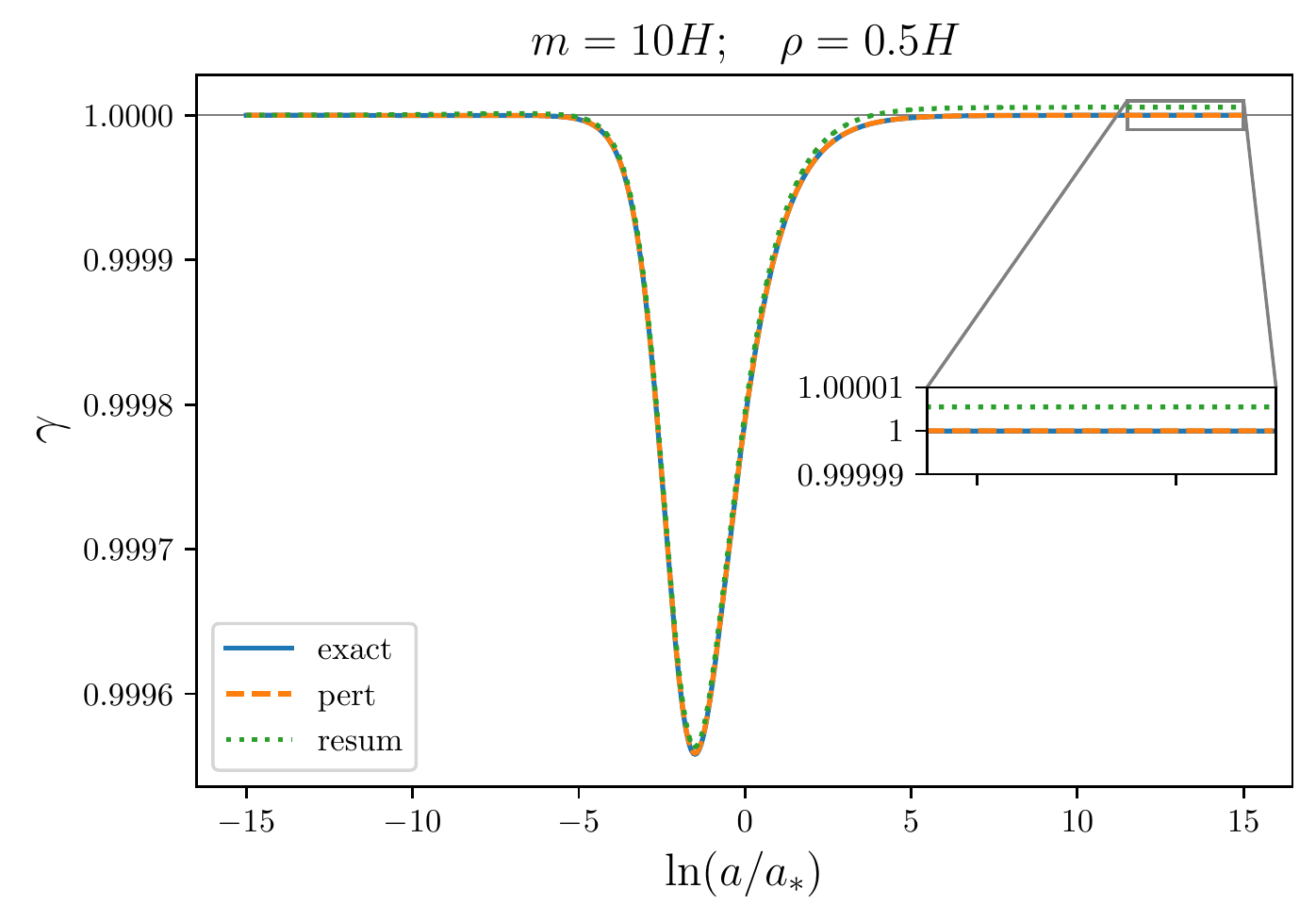}
\vspace{-0.8cm}
\caption{Same as in \Fig{fig:reco}, where the solution of the master equation~\eqref{eq:eq3} is also shown (green line), together with its perturbative limit (orange line).
The agreement is excellent, and becomes even better when decreasing $\rho$ or when increasing $m$.
At late time, the full master equation does not perform better than the perturbative theory, and even leads to slight violations of the positivity of the dynamical map (see the inset where $\gamma>1$). 
}
\vspace{-0.4cm}
\label{fig:fig3}
\end{figure}

Master equations are commonly employed in cosmology to model the effect of additional degrees of freedom, treated as an environment, onto a given system \cite{FEYNMAN1963118, CALDEIRA1983587, PhysRevD.45.2843, Hu:1993qa, Eisert:2001je, Burgess:2007pt, Anastopoulos:2013zya, Boyanovsky:2015tba, Boyanovsky:2015jen,Boyanovsky:2018soy, Hollowood:2017bil, Shandera:2017qkg, Banerjee:2021lqu, Brahma:2021mng, Brahma:2022yxu, Prudhoe:2022pte, Colas:2022hlq, Kaplanek:2022xrr, Burgess:2022nwu, Cao:2022kjn}. 
One of their appealing advantages is their ability to re-sum late-time secular effects \cite{Burgess:2014eoa, Boyanovsky:2015tba, Boyanovsky:2015jen, Burgess:2015ajz, Kaplanek:2021fnl, Chaykov:2022zro, Chaykov:2022pwd, Colas:2022hlq}, hence to go beyond standard perturbation theory and implement non-perturbative resummations in cosmology. 
Note that the recoherence phenomenon being a manifestly non-Markovian feature, it cannot be modelled by an irreversible dynamical-map such as the Lindblad equation~\cite{Lindblad:1975ef}. 
It requires the use of more sophisticated non-Markovian master equations such as the time-convolutionless (TCL) master equation discussed in \Refs{1974Phy....74..215V, 1974Phy....74..239V, 2002quant.ph..9153B, Breuer:2002pc, Colas:2022hlq},
\begin{align}\label{eq:eq3}
\frac{\dd \widehat{\rho}_{\mathrm{red}}}{\dd \eta}  &=-i\left[	\widehat{H}^{\mathcal{S}}_0(\eta) + \widehat{H}^{\mathrm{(LS)}}(\eta),\widehat{\rho}_{\mathrm{red}}(\eta)\right] \\
+& \bs{\mathcal{D}}_{ij}(\eta) \left[\widehat{\bs{z}}_{\zeta,i}\widehat{\rho}_{\mathrm{red}}(\eta) \widehat{\bs{z}}_{\zeta,j}-\frac{1}{2}\left\{\widehat{\bs{z}}_{\zeta,j}\widehat{\bs{z}}_{\zeta,i},\widehat{\rho}_{\mathrm{red}}(\eta)\right\}\right] .  \nonumber
\end{align}
Here, $\widehat{H}^{\mathcal{S}}_0$ is the free Hamiltonian of the system, and the effect of the environment is encoded into the ``Lamb-shift'' Hamiltonian $\widehat{H}^{\mathrm{(LS)}}$ and the dissipator matrix $\bs{\mathcal{D}}$.
These objects are constructed out of the two-point functions of the environment and formally rely on convolutions from initial time to final time of the free mode functions of the system and the environment. 
Their detailed expression is obtained following the procedure of \Refa{Colas:2022hlq} in the SM, where the master equation~\eqref{eq:eq3} is derived explicitly.
The Lamb-shift term captures the renormalisation of the free Hamiltonian due to the interactions with the environment.
At late-time where $k \ll aH$, it yields an effective speed of sound $\cs^{2} = 1 - \rho^2/m^2 + \mathcal{O}[k/(aH),H^4/m^4]$, which rescales the kinetic term by $\widehat{p}_{\zeta}^2 \rightarrow \cs^{2}\widehat{p}_{\zeta}^2$. 
This effect is also found in Wilsonian EFT treatments of the model~\cite{Tolley:2009fg,Achucarro:2010da,Cespedes:2012hu, Achucarro:2012sm, Pi:2012gf}. Although non-perturbative, it only leads to a slight rescaling of the power spectra as mentioned above.  
This correction is however unitary, hence it cannot account for de(re)coherence~\cite{Breuer:2002pc}, which is instead driven by the second line of \Eq{eq:eq3}.
From \Eq{eq:eq3}, one can derive effective transport equations for $\bs{\Sigma}(\eta)$~\cite{Colas:2022hlq}, given in the SM.
This leads to the purity shown in \Fig{fig:fig3}, where ``resum'' stands for the full solution of \Eq{eq:eq3}, in which partial resummation is supposed to take place; and ``pert'' corresponds to the solution at leading order in $\rho^2$ [since $\bs{\mathcal{D}}=\mathcal{O}(\rho^2)$ this amounts to evaluating $\widehat{\rho}_{\mathrm{red}}$ in the free theory in the second line of \Eq{eq:eq3}]. 
In \Refa{Colas:2022hlq}, this was shown to coincide with the result of the in-in formalism~\cite{Weinberg:2006ac, Adshead:2009cb, Chen:2017ryl, Dimastrogiovanni:2022afr}. In practice, in the SM it also allows us to unambiguously identify and remove the so-called ``spurious terms'', which cancel out at leading order but otherwise spoil the resummation~\cite{Colas:2022hlq}.  
%
%
In \Fig{fig:fig3} one can see that the master equation provides an excellent fit to the full result, both in its perturbative limit and when solved entirely. 
In particular, it accurately captures the turning point of the purity. 
This is remarkable, since the highly non-Markovian nature of the recoherence phenomenon may have cast some doubts on the existence of an effective single-field description that would be under control.

One also notices that the non-perturbative resummation performed by the master equation does not significantly improve the perturbative treatment.
The reason is that, as stressed above, the coupling between the adiabatic and entropic sector is effectively switched off at late time. 
There is therefore no secular effects to be resummed, and the perturbative and non-perturbative results only differ by overall constant factors in the power spectra, as checked explicitly in the SM.
Therefore, the only non-perturbative effect is unitary. It consists in the rescaling of $\cs$ mentioned above, which only advances horizon crossing. 
This contrasts with the situation studied in \Refa{Colas:2022hlq}, where an effective mass is generated for the light degree of freedom. 
This dresses the anomalous dimension of the light field, generating a secular growth at the perturbative level which is then resummed by the master equation (or other methods such as the dynamical renormalisation group~\cite{Boyanovsky:1998aa,Burgess:2009bs, Green:2020txs}).

Finally, let us note that the master equation leads to a tiny violation of positivity at late time, see the inset in \Fig{fig:fig3} where the purity slightly overshoots one, implying that $\det[\bs{\Sigma}(\eta)] < 1/4$ (hence violating Heisenberg inequality \cite{Breuer:2002pc}).
This signals a small breakdown of the effective theory, and determining under which conditions this class of non-Markovian Gaussian dynamical maps remains completely positive and trace preserving (CPTP) would deserve further investigations~\cite{2008JPhA...41q5304W, Di_si_2014, Ferialdi_2016,RevModPhys.88.021002, Kaplanek:2022xrr}.

\vspace{0.2cm}
\noindent {\bf Conclusions.}
%
In this Letter, we have shown that heavy entropic degrees of freedom do \emph{not} lead to quantum decoherence of adiabatic fluctuations in the early universe, at least through their dominant interaction term. 
More precisely, we found that after a transient phase of decoherence, the adiabatic fluctuations recohere once the mode under consideration crosses out the Compton wavelength of the entropic field.
This is because, at late time, the interaction is effectively quenched off as a result of spacetime expansion. This makes the state purity freeze to a value close to unity. Therefore, heavy entropic fields leave a small imprint not only on cosmological observables, but also on quantum-information properties of the quantum state.

We also found that an effective master equation derived from open-quantum-system methods performs remarkably well when compared to the full theory.
Wilsonian EFTs have also been used to describe the model, but they do not capture non-unitary effects, hence they cannot describe decoherence.
The master equation treatment has allowed us to check that non-unitary effects are negligible in the observables of the system (hence Wilsonian EFTs can safely be used in that respect), although they are crucial as far as decoherence is concerned.
Let us stress that, since recoherence is inherently a  non-Markovian process, the master equation needs to be kept non-Markovian too, \ie beyond the Lindblad limit.
We noted that, due to the effective decoupling between adiabatic and entropic modes at late time, there is no secular growth that the master equation would otherwise resum.
This even leads to a slight violation of positivity by the effective dynamical map, which questions its ability to account for non-perturbative effects in the absence of secular divergences (when secular terms are present, non-perturbative resummation was found to be successful in \Refa{Colas:2022hlq}).

These results do not preclude other decoherence channels (such as higher-order coupling between adiabatic and entropic fluctuations, single-field gravitational decoherence~\cite{Burgess:2022nwu}, etc.) to effectively decohere cosmological perturbations, but it suggests that decoherence in the early universe may not be as ubiquitous as common wisdom suggests. 
This is crucial to determine whether or not genuine quantum signals can be detected in cosmological structures~\cite{Campo:2005sv, Martin:2016nrr, Martin:2017zxs, Kanno:2020usf, Berera:2021xqa, Martin:2021znx,  Martin:2022kph, Berera:2022nzs}.
Natural prospects of our work include the investigation of models with sharp turns \cite{Cespedes:2012hu, Raveendran:2023dst}, the impact of multiple entropic directions~\cite{Pinol:2021aun} on the emergence of Markovianity \cite{RevModPhys.88.021002}, as well as non-linear interactions~\cite{Pimentel:2022fsc, Jazayeri:2022kjy, Wang:2022eop}. 
In this latter case, mode coupling is expected to enlarge the size of the effective environment, but also to induce non-Gaussianities~\cite{Martin:2018lin}, which are tightly constrained~\cite{Planck:2019kim, DAmico:2022gki, Cabass:2022wjy, Riquelme:2022amo}.
One may also study how our results vary when changing the initial quantum state~\cite{Lesgourgues:1996jc,Baunach:2021yvu, Wenderoth:2022acy, Letey:2022hdp, Ghosh:2022cny}.

Let us end by stressing that, when the system is coupled to a single mode as in the present setting (and as in the two-field model of \Refa{Colas:2022hlq}), decoherence or recoherence are possible only because we work in a dynamical background. In flat spacetime indeed, as explained above finite-size environments and time-independent Hamiltonians can only lead to oscillations in the purity. This is a consequence of Poincar\'e recurrence-time theorem \cite{Poincare:1890}, which relies on volume conservation. In cosmology however, the large-scale dynamics either amplifies or extinguishes the effective coupling, which yields decoherence or recoherence respectively. Those phenomena have therefore no flat-space analogue. Since most open-quantum-system methods are developed in the context of laboratory experiments, hence in flat spacetimes, their use in cosmology requires a critical analysis of their applicability, to which this work hopefully contributes.

\vspace{0.5 cm}
\noindent {\it Acknowledgements.} 
It is a pleasure to thank Cliff Burgess, Richard Holman, Greg Kaplanek, Ancel Larzul, J\'er\^ome Martin, Amaury Micheli, S\'ebastien Renaux-Petel, Marco Schir\'o, Mattia Walschaers and Denis Werth for interesting discussions.

\bibliographystyle{apsrev4-2}
\bibliography{reco.bib}

\clearpage
\onecolumngrid
\begin{center}
\textbf{\large SUPPLEMENTAL MATERIAL \\[.2cm] ``Quantum recoherence in the early universe''}\\[.2cm]
\vspace{0.05in}
{Thomas Colas, Julien Grain, and Vincent Vennin}
\end{center}

\setcounter{equation}{0}
\setcounter{figure}{0}
\setcounter{table}{0}
\setcounter{section}{0}
\setcounter{page}{1}
\makeatletter
\renewcommand{\theequation}{S\arabic{equation}}
\renewcommand{\thefigure}{S\arabic{figure}}
\renewcommand{\thetable}{S\arabic{table}}

\onecolumngrid

This supplemental material contains some technical details of the calculations presented in the main text.
In \Sec{sec:exact}, we derive the transport equations whose solutions are given in the main text.
In \Sec{sec:Minkowski}, we show that purity exhibits oscillation when our setting is considered in flat spacetime.
In \Sec{sec:TCL}, we derive and solve the master equation associated to our problem, closely following \Refa{Colas:2022hlq}.
%

\section{Exact solution}
\label{sec:exact}

\subsection{Hamiltonian formulation}

Starting from the Lagrangian density 
\begin{align}
\mathcal{L} &=  a^2  \epsilon M_{\mathrm{Pl}}^2 \zeta^{\prime2} - a^2 \epsilon M_{\mathrm{Pl}}^2  \left(\partial_i \zeta\right)^2  
+ \frac{1}{2}a^2 \mathcal{F}^{\prime2} - \frac{1}{2}a^2 \left(\partial_i  \mathcal{F}\right)^2
- \frac{1}{2}m^2 a^4 \mathcal{F}^2 - \rho a^3 \sqrt{2\epsilon} 
 M_{\mathrm{Pl}} \zeta^{\prime} \mathcal{F},
\end{align}
we first introduce the rescaled Mukhanov-Sasaki like variables $v_{\zeta}(\eta,\bs{x}) \equiv - a(\eta) \sqrt{2\epsilon} M_{\mathrm{Pl}} \zeta(\eta, \bs{x}) $ and $v_{\mathcal{F}}(\eta,\bs{x}) \equiv a(\eta) \mathcal{F}(\eta, \bs{x})$.
One can then Fourier transform the fields
\begin{align}\label{eq:Fourier}
v_{\alpha}(\eta,\bs{k}) \equiv  \int_{\mathbb{R}^{3}} \frac{\mathrm{d}^3 \boldmathsymbol{x}}{(2\pi)^{3/2}} v_{\alpha}(\eta,\bs{x}) e^{-i\boldsymbol{k}.\bs{x}} , 
\end{align}	
for $\alpha = \zeta, \mathcal{F}$. The conjugate momenta are obtained from \Eq{eq:eq1} and read
\begin{align}
\label{eq:momenta:def}
p_{\zeta} = v'_{\zeta} - \frac{a'}{a} v_{\zeta} + \rho a v_{\mathcal{F}}
\quad\text{and}\quad
p_{\mathcal{F}} = v'_{\mathcal{F}} - \frac{a'}{a} v_{\mathcal{F}}\, .
\end{align}
A Legendre transform gives the Hamiltonian
\begin{eqnarray}\label{eq:Hamiltu}
H = \int_{\mathbb{R}^{3+}} \mathrm{d}^3\boldmathsymbol{k} \bs{z}^{\dag}\boldsymbol{H}(\eta)\bs{z}\, ,
\end{eqnarray}
where the phase-space variables have been arranged into the vector $\bs{z} \equiv (v_\zeta, p_\zeta, v_{\mathcal{F}}, p_{\mathcal{F}})^{\mathrm{T}}$ and $\boldsymbol{H}$ is a four-by-four matrix given by
\bea\label{eq:Hmat}
\boldsymbol{H}(\eta)=\left(\begin{array}{cc}
\boldsymbol{H}^{(\mathcal{S})} & \boldsymbol{V} \\
\boldsymbol{V}^{\mathrm{T}} & \boldsymbol{H}^{(\mathcal{E})}
\end{array}\right),
\eea
with
\begin{align}\label{eq:Hvarphimat}	
\boldsymbol{H}^{(\mathcal{S})}(\eta) = \begin{pmatrix}
k^2 &\frac{a'}{a} \\ 
\frac{a'}{a} & 1
\end{pmatrix} ,
\quad
\boldsymbol{H}^{(\mathcal{E})}(\eta) = \begin{pmatrix}
k^2 + \left(m^2 + \rho^2\right) a^2 & \frac{a'}{a} \\ 
\frac{a'}{a} & 1
\end{pmatrix} ,
\quad
\bs{V}(\eta) \equiv \begin{pmatrix}
0 & 0 \\ 
-\rho a & 0
\end{pmatrix}. 
\end{align}
Note that, since $\zeta$ and $\mathcal{F}$ are real fields, one has the constrain $\bs{z}^{*}(\eta,\bs{k}) = \bs{z}(\eta,-\bs{k})$. This explains why, in order to avoid double counting, the integral in \Eq{eq:Hamiltu} is performed over $\mathbb{R}^{3+}\equiv\mathbb{R}^2\times\mathbb{R}^+$. Remarkably, the linear mixing $\rho$ enters the definition of the entropic mass $m^2 + \rho^2$ when the problem is described in terms of canonical variables. From now on, we thus redefine $m^2 \rightarrow m^2 + \rho^2$.

Following the canonical quantisation prescription, field variables are promoted to quantum operators. In order to work with hermitian operators, we split the fields into real and imaginary components, that is 
\begin{align}
\widehat{\boldmathsymbol{z}} &= \frac{1}{\sqrt{2}}\left(\widehat{\boldmathsymbol{z}}^{\mathrm{R}} + i \widehat{\boldmathsymbol{z}}^{\mathrm{I}}\right) , 
\end{align}
such that $\widehat{\boldmathsymbol{z}}^s$ is Hermitian for $s = \mathrm{R}, \mathrm{I}$. These variables are canonical since $[\widehat{v}^{s}_{\alpha}(\bs{k}),\widehat{p}^{s\prime}_{\alpha'}(\bs{q}) ] = i \delta^3(\boldsymbol{k}-\bs{q})\delta_{\alpha,\alpha'}\delta_{s,s\prime}$. In this basis, the Hamiltonian takes the same form as in \Eq{eq:Hamiltu}, \ie 
\bea
\label{eq:Hamiltonian:k:s}
\widehat{H}(\eta) =\frac{1}{2}\sum_{s=\mathrm{R},\mathrm{I}} \int_{\mathbb{R}^{3+}} \mathrm{d}^3\boldmathsymbol{k} \left(\widehat{\bs{z}}^s\right)^{\mathrm{T}}\boldsymbol{H}(\eta)\widehat{\bs{z}}^s\, .
\eea
Being separable, there is no mode coupling nor interactions between the $\mathrm{R}$ and $\mathrm{I}$ sectors and the state is factorisable in this decomposition. Hence, from now on, we focus on a given wavenumber $\bs{k}$ and a given $s$-sector, and to make notations lighter we leave the $\bs{k}$ and $s$ dependence implicit. 

\subsection{Transport equations}\label{subsec:transport}

\begin{figure}
\includegraphics[width=0.8\textwidth]{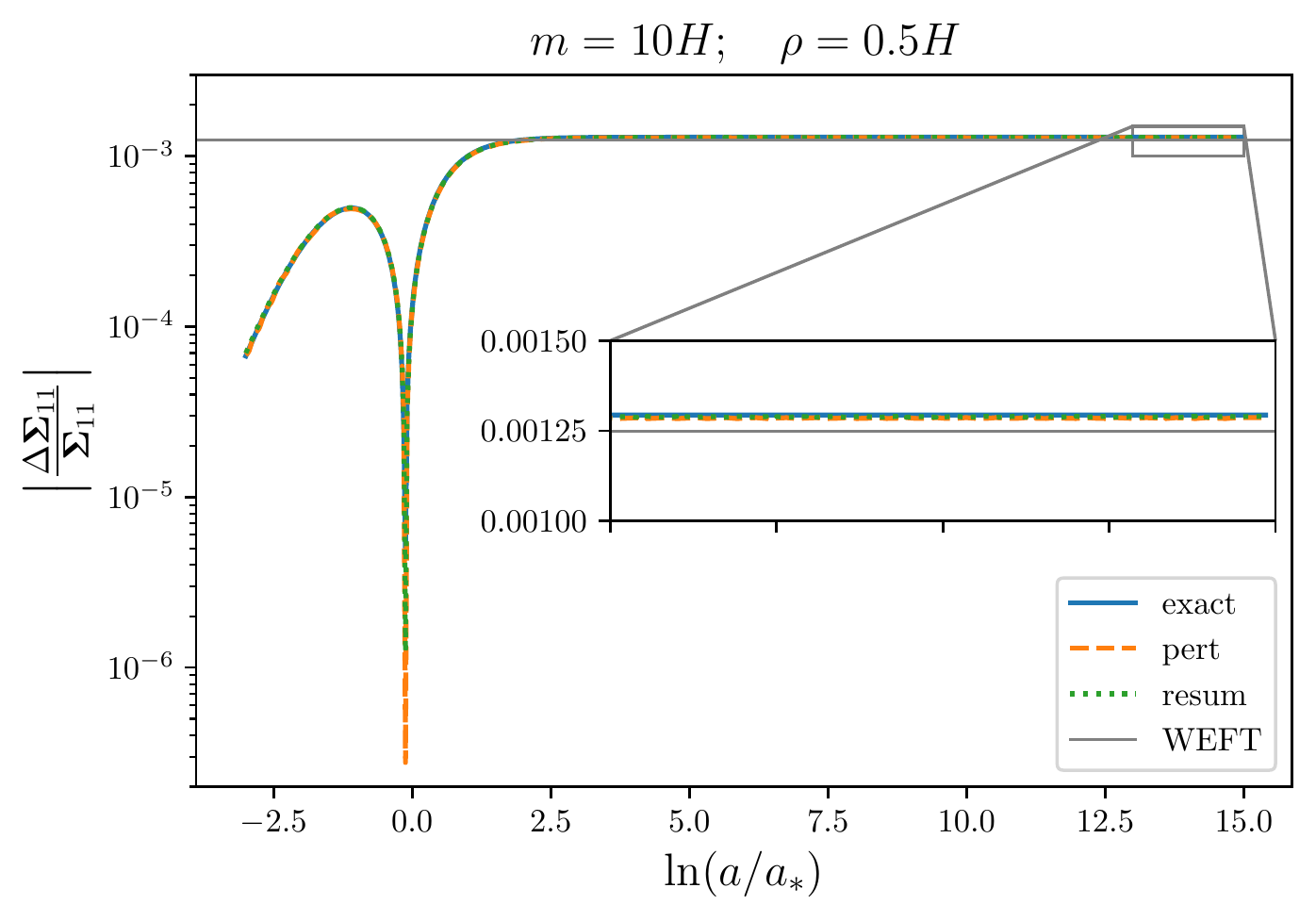}
\vspace{-0.4cm}
\caption{Relative correction to the free configuration-configuration power spectrum $\vert \bs{\Sigma}_{11}-\bs{\Sigma}^{(0)}_{11}\vert/\bs{\Sigma}^{(0)}_{11}$, as a function of time labeled by the number of \efolds~$\ln(a/a_*)$ since Hubble crossing, for $m=10H$ and $\rho=0.5H$. The blue curve is obtained from integrating the exact transport equations~\eqref{eq:eomexact} between $\ln(a/a_*)=-15$ to $\ln(a/a_*)=15$. 
The green curve corresponds to the master transport equation~\eqref{eq:TCLeom}, and the orange curve to its perturbative limit~\eqref{eq:TCLeompert}. The grey curve stands for the late-time result from Wilsonian EFT~\cite{Achucarro:2010da, Cespedes:2012hu}, which reduces to the in-in formalism~\cite{Chen:2012ge,Pi:2012gf} in the perturbative limit. The slight deviation from the orange curve is due to the additional expansion in $H/m$ usually performed in WEFT. 
Similar behaviours are observed for the other two power spectra, namely $\bs{\Sigma}_{12}$ and $\bs{\Sigma}_{22}$. }
\label{fig:spectra}
\vspace{-0.4cm}
\end{figure}

The transport equations for the full system-plus-environment setup can be obtained by differentiating 
\begin{align}
    \boldmathsymbol{\Sigma}^{(\mathcal{S}+\mathcal{E} )}_{ij}(\eta) \equiv \frac{1}{2} \Tr \left\{ [ \boldmathsymbol{\widehat{z}}_{i}(\eta)  \boldmathsymbol{\widehat{z}}_{j}(\eta) + \boldmathsymbol{\widehat{z}}_{j}(\eta)  \boldmathsymbol{\widehat{z}}_{i}(\eta) ] \widehat{\rho}_{0}\right\}
\end{align} 
with respect to time in the Heisenberg picture, and using the Heisenberg equations to evaluate $\dd\widehat{\bs{z}}/\dd\eta$. The density matrix $\widehat{\rho}_{0}$ specifies the initial Bunch-Davies vacuum in which the adiabatic and entropic directions both start. The Hamiltonian~\eqref{eq:Hamiltonian:k:s} being quadratic, one finds
\bea \label{eq:eomexact}
\frac{\dd\boldsymbol{\Sigma}^{(\mathcal{S}+\mathcal{E} )}}{\dd \eta}=\boldsymbol{\Omega H \Sigma}^{(\mathcal{S}+\mathcal{E} )}-\boldsymbol{\Sigma}^{(\mathcal{S}+\mathcal{E} )} \boldsymbol{H \Omega},
\eea
where $\bs{H}$ was defined in \Eq{eq:Hmat} and $\boldsymbol{\Omega}$ is a four-by-four block-diagonal matrix where each $2\times2$ block on the diagonal is the symplectic matrix $\bs{\omega} = \begin{pmatrix}
    0 & 1 \\
    -1 & 0
\end{pmatrix}$. Once \Eq{eq:eomexact} is known, one can derive an exact equation for $\det  \bs{\Sigma}$ 
\begin{align}\label{eq:detexact}
\frac{\dd\det \bs{\Sigma}}{\dd\eta} = \boldsymbol{\Sigma}^{(\mathcal{S}+\mathcal{E} )}_{11}  \frac{\dd\boldsymbol{\Sigma}^{(\mathcal{S}+\mathcal{E} )}_{22} }{\dd\eta} + \boldsymbol{\Sigma}^{(\mathcal{S}+\mathcal{E} )}_{22}  \frac{\dd\boldsymbol{\Sigma}^{(\mathcal{S}+\mathcal{E} )}_{11} }{\dd\eta} - 2 \boldsymbol{\Sigma}^{(\mathcal{S}+\mathcal{E} )}_{12}  \frac{\dd\boldsymbol{\Sigma}^{(\mathcal{S}+\mathcal{E} )}_{12} }{\dd\eta}
\end{align}
where $ \bs{\Sigma}$ is the system's covariance. \Eqs{eq:eomexact} and \eqref{eq:detexact} provide a set of eleven coupled ordinary differential equations that we numerically integrate from $\log (-k\eta_{\mathrm{ini}} )= 15$ to $\log(-k\eta_{\mathrm{fin}}) = -15$. Note that \Eq{eq:detexact} is redundant with \Eqs{eq:eomexact}, but since it arises from cancellations between quantities that diverge at late time, to compute the purity it is numerically more efficient to treat it as independent. Initial conditions are computed in the Bunch-Davies vacuum where $\boldsymbol{\Sigma}^{(\mathcal{S}+\mathcal{E} )}_{11} = \boldsymbol{\Sigma}^{(\mathcal{S}+\mathcal{E} )}_{33}  = 1/(2k)$, $\boldsymbol{\Sigma}^{(\mathcal{S}+\mathcal{E} )}_{22} = \boldsymbol{\Sigma}^{(\mathcal{S}+\mathcal{E} )}_{44}  = k/2$ and all other correlations initially vanish. The entries of $ \bs{\Sigma}$ are shown in \Fig{fig:spectra} where we observe that, at late-time, the effect of the heavy environment is to simply rescale the amplitude of the power spectra by approximately $\rho^2/(2m^2)$, in agreement with the results derived in the past literature with effective methods, see e.g. \Refs{Chen:2012ge,Pi:2012gf} for the in-in treatment and \Refs{Achucarro:2010da, Cespedes:2012hu} for the Wilsonian EFT approach. 

\subsection{Mode functions}\label{subsec:mode}

In the interaction picture, operators evolve according to the free Hamiltonian $\widehat{H}_0(\eta)= \widehat{H}^{\mathcal{S}}_0(\eta) \otimes \widehat{H}^{\mathcal{E}}_0(\eta)$ where
\begin{align}
\widehat{H}^{\mathcal{S}}_0(\eta) &= \frac{1}{2} \left(\widehat{p}_{\zeta}\widehat{p}_{\zeta} + k^2 \widehat{v}_{\zeta}\widehat{v}_{\zeta} + \frac{a'}{a} \left\{\widehat{v}_{\zeta}, \widehat{p}_{\zeta} \right\}\right) \\ 
\widehat{H}^{\mathcal{E}}_0(\eta) &= \frac{1}{2} \left[\widehat{p}_{\mathcal{F}}\widehat{p}_{\mathcal{F}} + \left(k^2 + m^2 a^2\right) \widehat{v}_{\mathcal{F}}\widehat{v}_{\mathcal{F}} + \frac{a'}{a} \left\{\widehat{v}_{\mathcal{F}}, \widehat{p}_{\mathcal{F}} \right\} \right]
\end{align}
with $\{A,B\} = AB+BA$ the anticommutator, while states and density matrices evolve according to the interaction Hamiltonian 
\begin{align}\label{eq:Hint}
\widehat{H}_{\mathrm{int}}(\eta) &= - \rho a(\eta) \widehat{p}_{\zeta} \widehat{v}_{\mathcal{F}}
\end{align}
where we used the fact that the $\zeta$ and $\mathcal{F}$ sectors commute. In this picture, the field operators admit a simple mode-function decomposition
\begin{align}\label{eq:modefctdecomp}
\widetilde{v}_{\alpha}(\eta) = v_{\alpha}(\eta) \widehat{a}_{\alpha} +  v^{*}_{\alpha}(\eta) \widehat{a}^{\dag}_{\alpha}
\end{align}
where $\widehat{a}_{\alpha}$ and $\widehat{a}^{\dag}_{\alpha}$ are the creation and annihilation operators of the uncoupled fields. From now on, tildas denote operators in the interaction picture. Heisenberg's equations yield  the classical equations of motion for the mode functions, \ie
\begin{eqnarray}
\label{eq:dyn1} v''_{\zeta} + \left( k^2 -  \frac{2}{\eta^2}\right)v_{\zeta} = 0
\quad\text{and}\quad
\label{eq:dyn2} v''_{\mathcal{F}} + \left( k^2 -  \frac{\nu^2_{\mathcal{F}}-\frac{1}{4}}{\eta^2}\right)v_{\mathcal{F}} = 0\, .
\end{eqnarray}
In these expressions, $\nu_{\mathcal{F}} = \frac{3}{2}\sqrt{1 - \frac{4}{9} \frac{m^2}{H^2}} \equiv i \mu_{\mathcal{F}}$ if $m^2 > \frac{9}{4}H^2$, which we will assume to be the case in the following, except explicitly stated otherwise. By normalising the mode functions to the Bunch-Davies vacuum in the asymptotic, sub-Hubble past, one obtains
\begin{align}
\label{eq:modefctvp}	v_{\zeta}(\eta) &= - \frac{1}{2}\sqrt{\frac{\pi z}{k}}  H_{3/2}^{(1)}(z) = \left(1 + \frac{i}{z}\right)\frac{\ee^{iz}}{\sqrt{2k}}, \\
\label{eq:modefctvs}	v_{\mathcal{F}}(\eta) &=\frac{1}{2}\sqrt{\frac{\pi z}{k}} \ee^{-\frac{\pi}{2}\mu_{\mathcal{F}}+i\frac{\pi}{4}} H_{i\mu_{\mathcal{F}}}^{(1)}(z) \, .
\end{align}
In these expressions, $z \equiv - k \eta$ and $H^{(1)}_{\nu}$ is the Hankel function of the first kind and of order $\nu$. The mode functions of the momentum operators read 
\begin{align}
\label{eq:modefctpp}	p_{\zeta}(\eta) &= \frac{1}{2} \sqrt{k \pi z} H_{1/2}^{(1)}(z) = - i \sqrt{\frac{k}{2}} \ee^{iz}, \\
\label{eq:modefctps}	p_{\mathcal{F}}(\eta) &=-\frac{1}{2}\sqrt{\frac{k \pi}{z}}\ee^{-\frac{\pi}{2}\mu_{\mathcal{F}}+i\frac{\pi}{4}}\left[ \left(i\mu_{\mathcal{F}}+\frac{3}{2}\right) H_{i\mu_{\mathcal{F}}}^{(1)}(z)-zH_{i\mu_{\mathcal{F}}+1}^{(1)}(z) \right],
\end{align}
where one can check that the mode functions are normalised in a way that the field operators obey their canonical commutation relations.

\section{Purity oscillations in Minkowski spacetime}
\label{sec:Minkowski}

\begin{figure}
\includegraphics[width=0.8\textwidth]{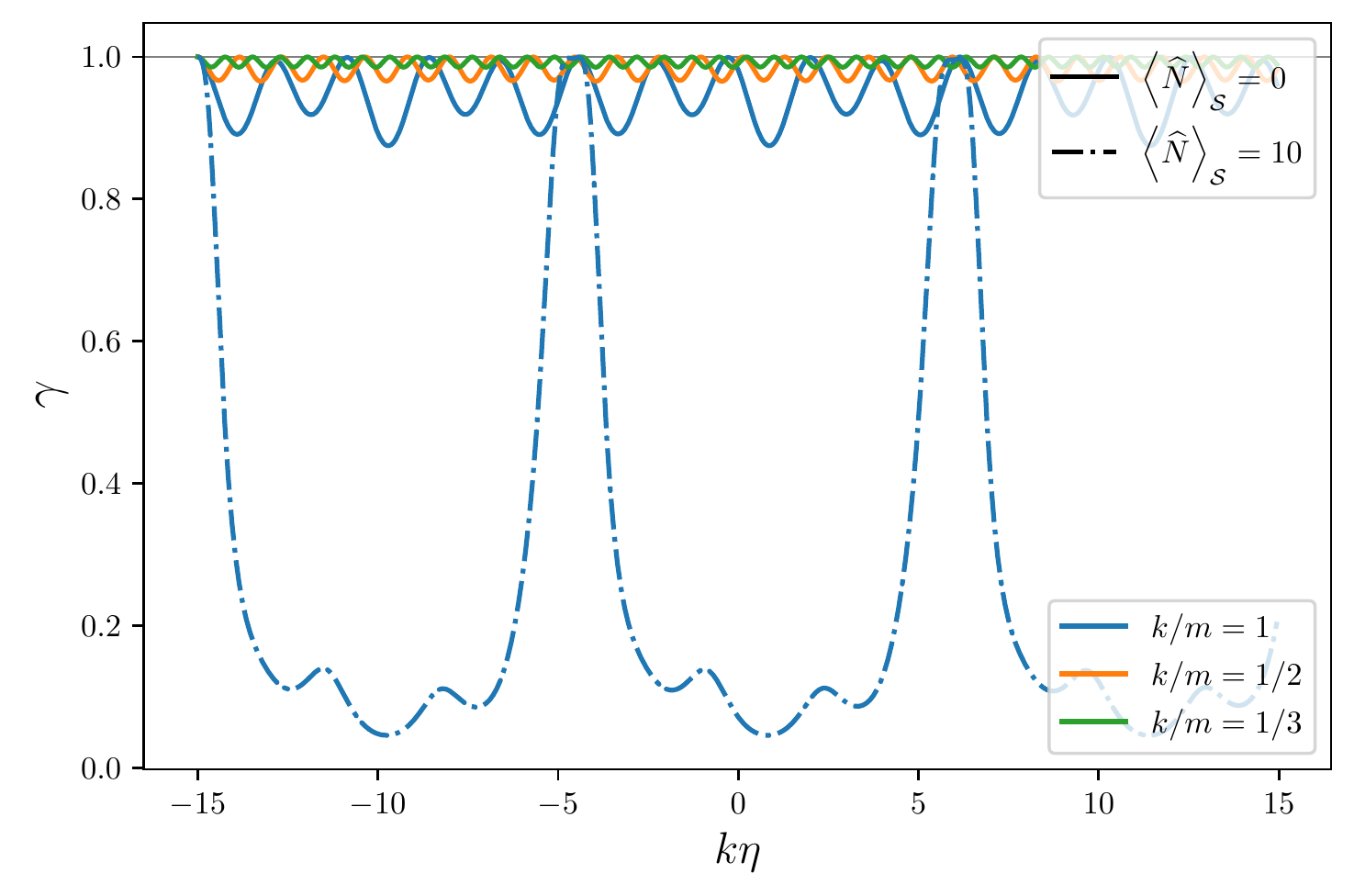}
\vspace{-0.4cm}
\caption{Purity in flat spacetime, as a function of time, for a few values of the ratio between the Compton wavelength and the physical wavelength $k/m$. Oscillations take place at frequencies $2\omega_{\mathcal{S}}$, $2\omega_{\mathcal{E}}$, $\omega_{\mathcal{S}}+\omega_{\mathcal{E}}$ and $\omega_{\mathcal{S}}-\omega_{\mathcal{E}}$; where $\omega_{\mathcal{S}} \equiv k$ and $\omega_{\mathcal{E}}\equiv \sqrt{k^2 + m^2}$. In the sub-Compton regime, $k/m > 1$,  the slowest frequency is $\omega_{\mathcal{E}}-\omega_{\mathcal{S}}$, which decreases with $k/m$ (this is why oscillations are more rapid for smaller values of $k/m$ in the figure). One can see that the amplitude of the oscillations also decreases as $k/m$ becomes smaller, in agreement with the fact that heavier environments yield weaker perturbations of the system.  Different initial states specified by $\left|\cancel{0} \right>_{\mathcal{S}}\otimes \left|\cancel{0} \right>_{\mathcal{E}} $ (solid curves) and $\left|\mathrm{2MSS}\right>_{\mathcal{S}}\otimes \left|\cancel{0} \right>_{\mathcal{E}} $ (dash-dotted curve) show that initial conditions also affect the system-environment entanglement but do not alter the recurrence phenomenon (\ie the fact that purity goes back to one, periodically), which is unavoidable.
}
\label{fig:Mink}
\vspace{-0.4cm}
\end{figure}

The flat-space analogue of the model studied in this work is obtained by taking the limit where $a=1$ and $a'=0$ in \Eq{eq:Hvarphimat}. This leads to
\Fig{fig:Mink} where we consider two sets of initial conditions. The first set consists in a vacuum state $\left|\cancel{0} \right>_{\mathcal{S}}\otimes \left|\cancel{0} \right>_{\mathcal{E}} $. For the initial covariance matrix, it gives the same prescription as above, $\boldsymbol{\Sigma}^{(\mathcal{S}+\mathcal{E} )}_{11} = \boldsymbol{\Sigma}^{(\mathcal{S}+\mathcal{E} )}_{33}  = 1/(2k)$, $\boldsymbol{\Sigma}^{(\mathcal{S}+\mathcal{E} )}_{22} = \boldsymbol{\Sigma}^{(\mathcal{S}+\mathcal{E} )}_{44}  = k/2$ where all other correlations initially vanish. The second set consists in a Gaussian state  $\left|\mathrm{2MSS}\right>_{\mathcal{S}}\otimes \left|\cancel{0} \right>_{\mathcal{E}} $ with $\left|\mathrm{2MSS}\right>_{\mathcal{S}}$ being a two-mode squeezed state chosen so that the initial occupation number of the system is
\begin{align}
    \left< \widehat{N} \right>_{\mathcal{S}} \equiv {}_{\mathcal{S}}{\left<\mathrm{2MSS}\right|} \widehat{a}^{\dag}_{\zeta}  \widehat{a}_{\zeta}\left|\mathrm{2MSS}\right>_{\mathcal{S}} = 10.
\end{align}
It amounts to picking a set of squeezing parameters~\cite{Colas:2021llj} $(r, \varphi)$ such that $\cosh r =2 \left< \widehat{N} \right>_{\mathcal{S}} + 1 $ and $\varphi$ is arbitrary, e.g. taken to zero, which fixes the initial correlations at 
\begin{align}
    \boldsymbol{\Sigma}^{(\mathcal{S}+\mathcal{E} )}_{11} &= \frac{1}{2k} \left(\cosh 2r + \sinh2r \cos{2\varphi}\right) \\
    \boldsymbol{\Sigma}^{(\mathcal{S}+\mathcal{E} )}_{22} &= \frac{k}{2} \left(\cosh 2r - \sinh2r \cos{2\varphi}\right) \\
    \boldsymbol{\Sigma}^{(\mathcal{S}+\mathcal{E} )}_{12} &= \frac{1}{2}  \sinh2r \sin{2\varphi}    
\end{align}
while keeping  $ \boldsymbol{\Sigma}^{(\mathcal{S}+\mathcal{E} )}_{33}  = 1/(2k)$, $\boldsymbol{\Sigma}^{(\mathcal{S}+\mathcal{E} )}_{44}  = k/2$ and all other correlations vanish. The evolution of the purity for both types of initial conditions in shown in \Fig{fig:Mink}, where we observe a recurrence phenomenon~\cite{Poincare:1890} at a frequency depending on the ratio between the Compton wavelength and the physical wavelength, $k/m$. Having a larger occupation number at initial time increases and fastens the system-environment entanglement but recurrence invariably occurs. 

\section{Master equation}
\label{sec:TCL}
%
In this section we review the approach developed in \Refa{Colas:2022hlq} and perform its direct application to the model considered in this work.

\subsection{Second-order master equation for a generic linear two-field systems}\label{subsec:TCL2}

Let us consider two scalar fields $\zeta$ (the ``system'') and $\mathcal{F}$ (the ``environment''), linearly coupled in a homogeneous and isotropic background. The coupling is assumed to be weak, which allows us to work within the Born-approximation regime. 
When expanding the dynamics of the system in powers of the coupling, at second order one obtains the time-convolutionless$_2$ (TCL${}_2$) master equation for the reduced density matrix of the system $\widetilde{\rho}_{\text{red}}$, which in the interaction picture reads
\begin{align}
\label{eq:TCL2exp}
\frac{\dd \widetilde{\rho}_{\text{red}}}{\dd \eta} = - \int_{\eta_0}^{\eta} \dd \eta' \Tr_{\mathcal{E}} \left[\widetilde{H}_{\text{int}}(\eta),\left[ \widetilde{H}_{\text{int}}(\eta'),  \widetilde{\rho}_{\text{red}}(\eta) \otimes \widetilde{\rho}_{\mathcal{E}}\right]\right]\, .
\end{align}
Here, the quadratic interaction Hamiltonian can be expressed as
\begin{eqnarray}
\widetilde{H}_{\text{int}}(\eta) =  \widetilde{\bs{z}}_{\zeta}^{\mathrm{T}}(\eta)\boldsymbol{V}(\eta)\widetilde{\bs{z}}_{\mathcal{F}}(\eta).
\end{eqnarray}
$\boldsymbol{V}(\eta)$ is an arbitrary $2\times2$ matrix containing the linear couplings between the two fields and  $\widetilde{\bs{z}}_{\alpha} = \left(\widetilde{v}_\alpha, \widetilde{p}_\alpha\right)^{\mathrm{T}}$, $\alpha = \zeta, \mathcal{F}$ gathers the configuration and momentum operators of the system and the environment. In order to write \Eq{eq:TCL2exp} in the Schr\"odinger picture, we need to recast it in terms of local-in-time operators for the system. We use the fact that in the interaction picture, operators evolve with the free Hamiltonian $\widehat{\calH}_0({\eta})$ so that 
\begin{align}
\widetilde{\bs{z}}_{\zeta}  (\eta') =& \bar{\mathcal{T}} \exp\left[{i \int_{\eta}^{\eta'} \widehat{\calH}_0({\eta}'')\dd{\eta}''}\right] \widetilde{\bs{z}}_{\zeta}  (\eta) \mathcal{T} \exp\left[{-i \int_{\eta}^{\eta'} \widehat{\calH}_0({\eta}'')\dd{\eta}''}\right] \\
=& \bs{G}^{(\mathcal{S})}(\eta',\eta)	\widetilde{\bs{z}}_{\zeta}  (\eta) \label{eq:Green}
\end{align}
where $\bs{G}^{(\mathcal{S})}(\eta',\eta) \equiv \mathrm{Tr}\left\{\left[\widehat{\bs{z}}^{\mathrm{T}}_{\zeta}(\eta'), \widetilde{\bs{z}}_{\zeta}(\eta)\right] \widehat{\rho}_{\mathcal{S}} \right\} $ is the Green's matrix of the free system, with $\widehat{\rho}_{\mathcal{S}}$ the initial state of the system. Developing \Eq{eq:TCL2exp} and expressing it in terms of equal-time operators using \Eq{eq:Green}, one finds
\begin{align}\label{eq:CME}
\frac{\dd \widetilde{\rho}_{\mathrm{red}}}{\dd \eta} =& - \int_{\eta_0}^{\eta}\dd \eta' 
\Big\{\left[\widetilde{\bs{z}}_{\zeta,i}  (\eta) \widetilde{\bs{z}}_{\zeta,j} (\eta) \widetilde{\rho}_{\mathrm{red}}(\eta) -  \widetilde{\bs{z}}_{\zeta,j}  (\eta) \widetilde{\rho}_{\mathrm{red}}(\eta)\widetilde{\bs{z}}_{\zeta,i}  (\eta) \right]\bs{\mathcal{D}}_{ij}^{>}(\eta,\eta') \nonumber\\
-&\left[\widetilde{\bs{z}}_{\zeta,i} (\eta)  \widetilde{\rho}_{\mathrm{red}}(\eta) \widetilde{\bs{z}}_{\zeta,j}  (\eta) -   \widetilde{\rho}_{\mathrm{red}}(\eta) \widetilde{\bs{z}}_{\zeta,j} (\eta) \widetilde{\bs{z}}_{\zeta,i} (\eta) \right]\bs{\mathcal{D}}_{ij}^{>*}(\eta,\eta') 
\Big\} ,
\end{align} 
where implicit summation over repeated indices apply. The memory kernel $\bs{\mathcal{D}}^{>}(\eta,\eta')$ is defined by 
\begin{align}\label{eq:memory}
\bs{\mathcal{D}}^{>}(\eta,\eta') \equiv \bs{V}(\eta)\bs{\mathcal{K}}^{>}(\eta,\eta')\bs{V}^{\mathrm{T}}(\eta')\bs{G}^{(\mathcal{S})}(\eta',\eta)
\end{align}
where $\bs{\mathcal{K}}^{>}(\eta,\eta') \equiv \mathrm{Tr}\left[\widehat{\bs{z}}^{\mathrm{T}}_{\mathcal{F}}(\eta) \widetilde{\bs{z}}_{\mathcal{F}}(\eta') \widehat{\rho}_{\mathcal{E}}\right]$ is the Wightman function of the free environment with $\widetilde{\rho}_{\mathcal{E}}$ the initial state of the environment. One can finally decompose the memory kernel in real and imaginary parts $\bs{\mathcal{D}}^{>}(\eta,\eta') \equiv \bs{\mathcal{D}}^{\mathrm{Re}}(\eta,\eta') + i\bs{\mathcal{D}}^{\mathrm{Im}}(\eta,\eta') $. After some straightforward manipulations, one obtains the $\text{TCL}_2$ master equation in the Schr\"odinger picture
\begin{align}
\frac{\dd \widehat{\rho}_{\mathrm{red}}}{\dd \eta}  &=-i\left[\widehat{H}^{(\mathcal{S})}(\eta) + \widehat{H}^{\mathrm{(LS)}}(\eta),\widehat{\rho}_{\mathrm{red}}(\eta)\right] + \left[\bs{D}_{ij}(\eta) - i \Delta_{-}(\eta) \bs{\omega}_{ij}\right]\left[\widehat{\bs{z}}_{\zeta,i}\widehat{\rho}_{\mathrm{red}}(\eta) \widehat{\bs{z}}_{\zeta,j}-\frac{1}{2}\left\{\widehat{\bs{z}}_{\zeta,j}\widehat{\bs{z}}_{\zeta,i},\widehat{\rho}_{\mathrm{red}}(\eta)\right\}\right] .
\end{align}
The Lamb-shift Hamiltonian is a quadratic form $\widehat{H}^{\mathrm{(LS)}}(\eta) = \frac{1}{2} \widehat{\bs{z}}^{\mathrm{T}}_{\zeta} \bs{\Delta}(\eta) \widehat{\bs{z}}_{\zeta}$ where 
\begin{align}
\bs{\Delta}_{ij}(\eta)   = 2 \int_{\eta_0}^{\eta} \dd \eta' \bs{\mathcal{D}}^{\mathrm{Im}}_{(ij)}(\eta,\eta').
\end{align}
The noise and dissipation kernels are respectively defined as
\begin{align}
\bs{D}_{ij}(\eta) &= 2 \int_{\eta_0}^{\eta} \dd \eta'  \bs{\mathcal{D}}^{\mathrm{Re}}_{(ij)}(\eta,\eta')  \\ 
\Delta_{-}(\eta) &= 2 \int_{\eta_0}^{\eta} \dd \eta' \bs{\mathcal{D}}^{\mathrm{Im}}_{-}(\eta,\eta')
\end{align}
 where we used the symmetric and antisymmetric decomposition of $2\times2$ matrices $\bs{A}_{ij} = \bs{A}_{(ij)} + A_{-} \bs{\omega}_{ij}$ where $\bs{A}_{(ji)} = \bs{A}_{(ij)}$. For the specific model discussed in this article, the definition of $\boldsymbol{V}(\eta)$ is given in \Eq{eq:Hvarphimat}. 

\subsection{Cosmological master equation}

Using the mode-function decomposition of the fields obtained in \Sec{subsec:mode}, we derive the Wightman function of the environment
\begin{align}\label{eq:Wightman}
\bs{\mathcal{K}}^{>}(\eta,\eta') = \begin{pmatrix}
v_{\mathcal{F}}(\eta)v^{*}_{\mathcal{F}}(\eta') & p_{\mathcal{F}}(\eta)v^{*}_{\mathcal{F}}(\eta') \\
v_{\mathcal{F}}(\eta)p^{*}_{\mathcal{F}}(\eta') & p_{\mathcal{F}}(\eta)p^{*}_{\mathcal{F}}(\eta') 
\end{pmatrix},
\end{align}
and the Green's matrix of the system
\begin{align}\label{eq:Greenfin}
\bs{G}^{(\mathcal{S})}(\eta',\eta)	= 2 \begin{pmatrix}
-\Imag{p_{\zeta} (\eta) v^{*}_{\zeta} (\eta')} & \Imag{v_{\zeta} (\eta) v^{*}_{\zeta} (\eta')} \\
& \\
-\Imag{p_{\zeta} (\eta) p^{*}_{\zeta} (\eta')} & \Imag{v_{\zeta} (\eta) p^{*}_{\zeta} (\eta')}
\end{pmatrix},
\end{align}
where we used the Bunch-Davies initial vacuum prescription. Inserting \Eqs{eq:Wightman} and \eqref{eq:Greenfin} into the expression of the memory kernel given in \Eq{eq:memory}, we obtain the master equation presented in the main text which we rewrite here for convenience
\begin{align}
\frac{\dd \widehat{\rho}_{\mathrm{red}}}{\dd \eta}  &=-i\left[	\widehat{H}^{\mathcal{S}}_0(\eta) + \widehat{H}^{\mathrm{(LS)}}(\eta),\widehat{\rho}_{\mathrm{red}}(\eta)\right] + \bs{\mathcal{D}}_{ij}(\eta) \left[\widehat{\bs{z}}_{\zeta,i}\widehat{\rho}_{\mathrm{red}}(\eta) \widehat{\bs{z}}_{\zeta,j}-\frac{1}{2}\left\{\widehat{\bs{z}}_{\zeta,j}\widehat{\bs{z}}_{\zeta,i},\widehat{\rho}_{\mathrm{red}}(\eta)\right\}\right],  
\end{align}
where $\widehat{H}^{\mathrm{(LS)}}(\eta) = \frac{1}{2} \widehat{\bs{z}}^{\mathrm{T}}_{\zeta} \bs{\Delta}(\eta) \widehat{\bs{z}}_{\zeta}$ and $\bs{\mathcal{D}} \equiv \bs{D}(\eta) + i \Delta_{12}(\eta) \bs{\omega}$. The entries of the $\bs{\Delta}$ and $\bs{D}$ matrices are given by the so-called master-equation coefficients defined as
\begin{align}
\bs{\Delta}_{11}(\eta) &=0  \label{eq:Delta11}\\
\bs{\Delta}_{12}(\eta) &= \bs{\Delta}_{21}(\eta) =- 2 \rho^2 a(\eta) \int_{\eta_0}^{\eta} \dd \eta' a(\eta')\Imag{p_{\zeta} (\eta) p^{*}_{\zeta} (\eta')}\Imag{v_{\mathcal{F}} (\eta) v^{*}_{\mathcal{F}} (\eta') }  \label{eq:Delta12}\\ 
\bs{\Delta}_{22}(\eta) &= 4 \rho^2 a(\eta) \int_{\eta_0}^{\eta} \dd \eta' a(\eta')\Imag{v_{\zeta} (\eta) p^{*}_{\zeta} (\eta')}\Imag{v_{\mathcal{F}} (\eta) v^{*}_{\mathcal{F}} (\eta') },  \label{eq:Delta22}
\end{align}
and 
\begin{align}
\bs{D}_{11}(\eta) &=0 \label{eq:D11} \\
\bs{D}_{12}(\eta) &=\bs{D}_{21}(\eta) = - 2 \rho^2 a(\eta) \int_{\eta_0}^{\eta} \dd \eta' a(\eta')\Imag{p_{\zeta} (\eta) p^{*}_{\zeta} (\eta')}\Real{v_{\mathcal{F}} (\eta) v^{*}_{\mathcal{F}} (\eta') } \label{eq:D12}\\ 
\bs{D}_{22}(\eta) &= 4 \rho^2 a(\eta) \int_{\eta_0}^{\eta} \dd \eta' a(\eta')\Imag{v_{\zeta} (\eta) p^{*}_{\zeta} (\eta')}\Real{v_{\mathcal{F}} (\eta) v^{*}_{\mathcal{F}} (\eta') }. \label{eq:D22}
\end{align}

\subsection{Master-equation coefficients}\label{subsec:full}

A simple manipulation of \Eqs{eq:Delta12}, \eqref{eq:Delta22}, \eqref{eq:D12} and \eqref{eq:D22} leads to
\begin{align}
\bs{\Delta}_{12}(\eta)&=  -  \frac{\rho^2}{H^2} \frac{k}{z} \Real {p_{\zeta}(z)v_{\mathcal{F}}(z) \int_{z_0}^{z} \frac{\dd z'}{z'} p_{\zeta}^{*}(z') v_{\mathcal{F}}^{*}(z') - p_{\zeta}(z)v^{*}_{\mathcal{F}}(z) \int_{z_0}^{z} \frac{\dd z'}{z'} p_{\zeta}^{*}(z') v_{\mathcal{F}}(z')} \label{eq:Delta12bis}\\ 
\bs{\Delta}_{22}(z) &= 2  \frac{\rho^2}{H^2} \frac{k}{z} \Real{v_{\zeta}(z)v_{\mathcal{F}}(z) \int_{z_0}^{z} \frac{\dd z'}{z'} p_{\zeta}^{*}(z') v_{\mathcal{F}}^{*}(z') -v_{\zeta}(z)v^{*}_{\mathcal{F}}(z) \int_{z_0}^{z} \frac{\dd z'}{z'} p_{\zeta}^{*}(z') v_{\mathcal{F}}(z')},  \label{eq:Delta22bis}
\end{align}
and
\begin{align}
\boldsymbol{D}_{12}(\eta)&=   \frac{\rho^2}{H^2} \frac{k}{z} \Imag{p_{\zeta}(z)v_{\mathcal{F}}(z) \int_{z_0}^{z} \frac{\dd z'}{z'} p_{\zeta}^{*}(z') v_{\mathcal{F}}^{*}(z') + p_{\zeta}(z)v^{*}_{\mathcal{F}}(z) \int_{z_0}^{z} \frac{\dd z'}{z'} p_{\zeta}^{*}(z') v_{\mathcal{F}}(z')} \label{eq:D12bis}\\
\boldsymbol{D}_{22}(\eta) &= -2  \frac{\rho^2}{H^2} \frac{k}{z} \Imag{v_{\zeta}(z)v_{\mathcal{F}}(z) \int_{z_0}^{z} \frac{\dd z'}{z'} p_{\zeta}^{*}(z') v_{\mathcal{F}}^{*}(z')+ v_{\zeta}(z)v^{*}_{\mathcal{F}}(z) \int_{z_0}^{z} \frac{\dd z'}{z'} p_{\zeta}^{*}(z') v_{\mathcal{F}}(z') }, \label{eq:D22bis}
\end{align}
where we defined the variable $z \equiv - k \eta$. To obtain analytical expressions for the master-equation coefficients, we have to compute two integrals. The first one is 
\begin{align}
I_1(z,z_0) = \int_{z_0}^{z} \frac{\dd z'}{z'} p_{\zeta}^{*}(z') v_{\mathcal{F}}^{*}(z').
\end{align}
Inserting the mode function expressions given in \Eqs{eq:modefctvp}, \eqref{eq:modefctvs}, \eqref{eq:modefctpp} and \eqref{eq:modefctps}, we obtain
\begin{align}\label{eq:I1}
I_1(z,z_0) &=  \frac{i}{2} \sqrt{\frac{\pi}{2}} \ee^{-\frac{\pi}{2} \mu_{\mathcal{F}}} \ee^{-i \frac{\pi}{4}} \int_{z_0}^{z} \frac{\dd z'}{\sqrt{z'}}  \ee^{-iz'}H_{-i\mu_{\mathcal{F}}}^{(2)}(z') \equiv F_{I_1}(z) -  F_{I_1}(z_0) 
\end{align}
with
\begin{align}\label{eq:FI1}
F_{I_1}(z)	&= i \sqrt{\frac{\pi}{2}} \ee^{-\frac{\pi}{2} \mu_{\mathcal{F}}} \ee^{-i \frac{\pi}{4}}  \sqrt{z} \left[ \gamma^{*}_{\mu_{\mathcal{F}}}(z) g_{\mu_{\mathcal{F}}}( z) + \delta^{*}_{\mu_{\mathcal{F}}}(z) 	g_{-\mu_{\mathcal{F}}}(z)\right]
\end{align} 
where we have introduced for later convenience the notations
\begin{align}
\gamma_{\mu_{\mathcal{F}}}(z) &\equiv \frac{1+  \coth \pi \mu_{\mathcal{F}}}{\Gamma(1+i\mu_{\mathcal{F}})}\left(\frac{z}{2}\right)^{i\mu_{\mathcal{F}}},\quad\quad~~\,
\delta_{\mu_{\mathcal{F}}}(z) \equiv  \frac{-1}{\sinh \pi\mu_{\mathcal{F}}} \frac{1}{\Gamma(1-i\mu_{\mathcal{F}})}\left(\frac{z}{2}\right)^{ -i\mu_{\mathcal{F}} }
\end{align}
and
\begin{align}
g_{\mu_{\mathcal{F}}}( z)= \frac{1}{ 1 - 2 i \mu_{\mathcal{F}}}{{}_2F_2}^{\frac{1}{2} - i \mu_{\mathcal{F}},\frac{1}{2} - i \mu_{\mathcal{F}}}_{\frac{3}{2} - i \mu_{\mathcal{F}}, 1 - 2 i \mu_{\mathcal{F}}}(- 2iz),
\end{align}
${}_2 F_2 $ being the $(2,2)$ generalized hypergeometric function. Note that $	g^{*}_{\mu_{\mathcal{F}}}(z) = 	g_{-\mu_{\mathcal{F}}}(-z)$. The second integral is 
\begin{align}
I_2(z,z_0) = \int_{z_0}^{z} \frac{\dd z'}{z'} p_{\zeta}^{*}(z') v_{\mathcal{F}}(z')
\end{align}
and following the same procedure, one finds 
\begin{align}\label{eq:I2}
I_2(z,z_0) &= \frac{i}{2} \sqrt{-\frac{\pi}{2}} \ee^{-\frac{\pi}{2} \mu_{\mathcal{F}}} \ee^{i \frac{\pi}{4}} \int_{z_0}^{z} \frac{\dd z'}{\sqrt{z'}}  \ee^{-iz'}H_{i\mu_{\mathcal{F}}}^{(1)}(z') \equiv F_{I_2}(z) -  F_{I_2}(z_0) 
\end{align}
with
\begin{align}
F_{I_2}(z)	&= i \sqrt{\frac{\pi}{2}} \ee^{-\frac{\pi}{2} \mu_{\mathcal{F}}} \ee^{i \frac{\pi}{4}}  \sqrt{z} \left[ \delta_{\mu_{\mathcal{F}}}(z) g_{\mu_{\mathcal{F}}}(z)  + \gamma_{\mu_{\mathcal{F}}}(z) g_{-\mu_{\mathcal{F}}}(z) \right].
\end{align} 
Inserting \Eqs{eq:I1} and \eqref{eq:I2} into the expression of the master equation coefficients \eqref{eq:Delta12bis}, \eqref{eq:Delta22bis}, \eqref{eq:D12bis} and \eqref{eq:D22bis} and using the functions $F_{I_1}(z)$ and $F_{I_2}(z)$, we obtain analytic expressions for the $\text{TCL}_2$ coefficients. 

\subsubsection{Spurious terms}\label{subsec:spur}

In \Refa{Colas:2022hlq}, it has been shown that some terms dubbed ``spurious'' appear in the master-equation coefficients, that cancel out in the perturbative limit but ruin the resummation otherwise. More precisely, the master-equation coefficients are expressed as integrals between $\eta_0$ and $\eta$, see \Eqs{eq:Delta12bis}-\eqref{eq:D22bis}, i.e.
\bea
\label{eq:primF:def}
\bs{\Delta}_{12} = F_{\bs{\Delta}_{12}}\left(\eta,\eta\right)-F_{\bs{\Delta}_{12}}\left(\eta,\eta_0\right),
\eea 
where $ F_{\bs{\Delta}_{12}}(\eta,\cdot)$ is the primitive of the integrand appearing in \Eq{eq:Delta12bis}, which itself depends on $\eta$, and with similar notations for the other coefficients. The second term in \Eq{eq:primF:def}, the one that depends on the initial time $\eta_0$, is the ``spurious'' one. In the exact solution of \Sec{sec:exact}, there is no such initial-time dependent term in the dynamical equations, and indeed one can show that it cancels out at all orders in perturbation theory~\cite{Colas:2022hlq}. At leading order in the interaction strength, the master equation reduces to standard perturbation theory, hence again one can show that the spurious contribution vanishes~\cite{Colas:2022hlq}. At higher order however, the master equation stops being exact, since it only performs resummation of the leading-order interaction. This is why the spurious term alters the result. However, since we know that it should vanish at all orders, one can simply remove it by hand, and thus restore the ability of the master equation to perform efficient resummation~\cite{Colas:2022hlq}.  
One may be worried that, from \Eq{eq:primF:def}, the spurious terms are only defined up to an additive constant. However, since they are known to vanish at all (and in particular at leading) orders, they can be determined without ambiguity by comparison with the perturbative theory. In the following we thus remove spurious terms, which amounts to discarding all $ F_{I_1}(z_0) $ and $ F_{I_2}(z_0) $ terms in the above expressions.

\subsubsection{Super-Hubble limit}\label{subsec:SH}

We now exhibit the late-time super-Hubble limit of the master equation coefficients where we perform a systematic expansion in powers of $z \ll 1$. Expanding the mode functions and various elements appearing in \Eqs{eq:Delta12bis}, \eqref{eq:Delta22bis}, \eqref{eq:D12bis} and \eqref{eq:D22bis} in the super-Hubble regime, we obtain
\begin{align}
\bs{\Delta}_{12}(z)& =  \frac{\rho^2}{H^2}  \frac{16 k^2}{9+40\mu_{\mathcal{F}}^2+16\mu_{\mathcal{F}}^4}  \frac{z}{k}+ \mathcal{O}(z^3) \label{eq:Delta12SH}\\
\bs{\Delta}_{22}(z)&=  -\frac{\rho^2}{H^2}  \frac{1}{\frac{9}{4}+\mu_{\mathcal{F}}^2}+ \mathcal{O}(z^2), \label{eq:Delta22SH}
\end{align}
and
\begin{align}
\bs{D}_{12}(z)& =  -\frac{\rho^2}{H^2} \frac{1}{\mu_{\mathcal{F}}} \frac{(-6 + 8 \mu^2_{\mathcal{F}}) k^2}{9+40\mu_{\mathcal{F}}^2+16\mu_{\mathcal{F}}^4}  \frac{z}{k} + \mathcal{O}(z^3)\\
\bs{D}_{22}(z)&=  - \frac{3}{2}\frac{\rho^2}{H^2} \frac{1}{\mu_{\mathcal{F}}} \frac{1}{\frac{9}{4}+\mu_{\mathcal{F}}^2}+ \mathcal{O}(z^2).
\end{align}
Note that, in the heavy case where $\mu_{\mathcal{F}} \simeq m/H \gg 1$, \Eqs{eq:Delta12SH} and \eqref{eq:Delta22SH} lead to $\bs{\Delta}_{12} \ll a'/a $ and $\bs{\Delta}_{22} \rightarrow - \rho^2/m^2$, from which we deduce that the Lamb-shift Hamiltonian renormalises the free dynamics as 
\begin{align}
&\widehat{H}^{\mathcal{S}}_0(\eta) + \widehat{H}^{\mathrm{(LS)}}(\eta) \simeq \frac{1}{2} \left[\left(1- \frac{\rho^2}{m^2}\right)\widehat{p}_{\zeta}\widehat{p}_{\zeta} + k^2 \widehat{v}_{\zeta} \widehat{v}_{\zeta} +\frac{a'}{a}  \left\{\widehat{v}_{\zeta}, \widehat{p}_{\zeta}\right\} \right].
\end{align}
One can thus see that $\bs{\Delta}_{22}$ renormalises the kinetical term, generating an effective speed of sound
\begin{align}\label{eq:cs}
\cs^{2} = 1 - \frac{\rho^2}{m^2}+\mathcal{O}\left(\frac{k}{aH},\frac{H^4}{m^4}\right)
\end{align}
as stated in the main text.

\subsection{Effective transport equations}
\label{sec:transport}

\subsubsection{Transport equations derivation}

The covariance matrix of the system expressed in the Schr\"odinger picture reads
\begin{align}
    \label{eq:covdef}		\boldmathsymbol{\Sigma}_{ij}(\eta) \equiv \frac{1}{2} \Tr \left[ \left\{ \boldmathsymbol{\widehat{z}}_{\zeta,i}  , \boldmathsymbol{\widehat{z}}_{\zeta,j}  \right\} \widehat{\rho}_{\mathrm{red}}(\eta)\right].
\end{align}
By differentiating \Eq{eq:covdef} with respect to time and inserting \Eq{eq:eq3} in the right-hand side, we obtain the effective transport equations for the covariance matrix,
\begin{align}\label{eq:TCLeom}
\frac{\dd \bs{\Sigma}}{\dd \eta}&=\boldsymbol{\omega}\left(\bs{H}^{(\mathcal{S})}+\bs{\Delta}\right) \bs{\Sigma}-\bs{\Sigma} \left(\bs{H}^{(\mathcal{S})}+\bs{\Delta}\right) \boldsymbol{\omega}- \bs{\omega D \omega} +2\bs{\Delta}_{12}\bs{\Sigma}.
\end{align}
As mentioned above, a numerically efficient way to access the late-time behaviour of the purity is to derive an equation of motion for $\det\bs{\Sigma}$ from the transport equation of the covariance, leading to
\bea
\label{eq:TCLdet}
\frac{\dd\det\bs{\Sigma}}{\dd\eta}=\mathrm{Tr}\left(\bs{\Sigma} \bs{D} \right)
+4\bs{\Delta}_{12}\det\bs{\Sigma}.
\eea

\subsubsection{Perturbative treatment}
In the main text, the numerical solution of \Eqs{eq:TCLeom}-\eqref{eq:TCLdet} (labeled ``resum'' in Fig. 4 of the main text) is compared with a perturbative solution (labeled ``pert''), where the solution is derived at leading order in $\rho^2$. Since $\bs{\Delta}$ and $\bs{D}$ are of order $\rho^2$, this amounts to replacing $\bs{\Sigma}$ by its free-theory counterpart $\bs{\Sigma}^{(0)}$ when multiplied by  $\bs{\Delta}$ or $\bs{D}$ in the right-hand side of \Eqs{eq:TCLeom}-\eqref{eq:TCLdet}, 
\begin{align}\label{eq:TCLeompert}
\frac{\dd \bs{\Sigma}^{(2)}}{\dd \eta}&= \boldsymbol{\omega}\bs{H}^{(\mathcal{S})}\bs{\Sigma}^{(2)}-\bs{\Sigma}^{(2)}\bs{H}^{(\mathcal{S})} \boldsymbol{\omega} + \boldsymbol{\omega}\bs{\Delta} \bs{\Sigma}^{(0)}-\bs{\Sigma}^{(0)} \bs{\Delta} \boldsymbol{\omega} - \bs{\omega D \omega} +2\bs{\Delta}_{12}\bs{\Sigma}^{(0)}.
\end{align}
and
\bea
\label{eq:detSigma:eom}
\frac{\dd\det(\bs{\Sigma}^{(2)})}{\dd\eta}=\mathrm{Tr}\left(\bs{\Sigma}^{(0)}\bs{D}\right)
+\bs{\Delta}_{12},
\eea
where the superscript indicates the order at which a given observable is computed and we used the fact that $\det(\bs{\Sigma}^{(0)}) = 1/4$. In this limit, the environmental effects just play the role of source terms. In \Fig{fig:spectra}, the non-perturbative solution of \Eq{eq:TCLeom} and its perturbative limit~\eqref{eq:TCLeompert} are compared to the exact result. As explained in the main text, since the interaction is effectively switched off at late time, there is no substantial resummation in the current setting (contrary to the situation investigated in \Refa{Colas:2022hlq}). This is why the non-perturbative solution shows no sign of improvement at late time. 

\subsubsection{Super-Hubble expansion}

Inserting the super-Hubble expansion of the master equation coefficients obtained in \Sec{subsec:SH} into the transport equations~\eqref{eq:TCLeom}, and working order-by-order in $z$, one finds
\begin{align}
\bs{\Sigma}_{11}(z) =& A_{-2}^{\bs{\Sigma}_{11}}z^{-2} + f_1 \left(A_{-2}^{\bs{\Sigma}_{11}}\right) + f_2\left(A_{0}^{\bs{\Sigma}_{12}}\right)z\, , \label{eq:late11}\\
\bs{\Sigma}_{12}(z) =& -k A_{-2}^{\bs{\Sigma}_{11}}z^{-1} + A_{0}^{\bs{\Sigma}_{12}} + f_3\left(A_{0}^{\bs{\Sigma}_{12}}\right)z^2  \, ,\label{eq:late12}\\
\bs{\Sigma}_{22}(z) =& k^2 A_{-2}^{\bs{\Sigma}_{11}} -2k A_{0}^{\bs{\Sigma}_{12}}z + A_{2}^{\bs{\Sigma}_{22}} z^2 + 2 k f_3\left(A_{0}^{\bs{\Sigma}_{12}}\right) z^3\, .  \label{eq:late22} 
\end{align} 
Here, $A_{-2}^{\bs{\Sigma}_{11}}, A_{0}^{\bs{\Sigma}_{12}}$ and $A_{2}^{\bs{\Sigma}_{22}}$ are three constants that cannot be determined by a mere super-Hubble expansion, since they result from the full integrated dynamics (they can however be set by numerical matching to the full solution). In the free theory, they are given by $A_{-2}^{\bs{\Sigma}^{(0)}_{11}} = 1/(2k), A_{0}^{\bs{\Sigma}^{(0)}_{12}} = 0$ and $A_{2}^{\bs{\Sigma}^{(0)}_{22}} = 0$ but otherwise receive $\order{\rho^2}$ corrections.
We have also defined
\begin{align}
f_1 \left(A_{-2}^{\bs{\Sigma}_{11}}\right) &\equiv \left(1-\frac{\rho^2}{H^2}\frac{1}{\frac{1}{4}+\mu_{\mathcal{F}}^2} \right)A_{-2}^{\bs{\Sigma}_{11}} \\
f_2\left(A_{0}^{\bs{\Sigma}_{12}}\right) &\equiv - \frac{2}{3k} \left(1-\frac{\rho^2}{H^2}\frac{1}{\frac{9}{4}+\mu_{\mathcal{F}}^2} \right)A_{0}^{\bs{\Sigma}_{12}} + \frac{1}{2k} \frac{\rho^2}{H^2} \frac{1}{\mu_{\mathcal{F}}} \frac{1}{\frac{9}{4}+\mu_{\mathcal{F}}^2}  \\
f_3\left(A_{0}^{\bs{\Sigma}_{12}}\right) &\equiv \frac{2}{3}\left[1-\frac{\rho^2}{H^2}\left(\frac{1}{9+4\mu_{\mathcal{F}}^2} +\frac{3}{1+4\mu_{\mathcal{F}}^2}
 \right)\right] A_{0}^{\bs{\Sigma}_{12}} + \frac{\rho^2}{H^2}  \frac{4}{9+40\mu_{\mathcal{F}}^2+16\mu_{\mathcal{F}}^4}.
\end{align}
This shows that the presence of the environment does not change the dominant scaling in $z$, but simply modifies the coefficients of the expansion by $\order{\rho^2}$-suppressed corrections. This is again evidence that no secular growth needs to be resummed, and the main effect is perturbative. 

One can also use these results to extract the super-Hubble behaviour of the purity parameter. Using \Eqs{eq:late11}-\eqref{eq:late22}, and expanding in $\rho^2$, one finds
\bea
\det\bs{\Sigma}=\underbrace{\frac{1}{4}+\frac{A_{2}^{\bs{\Sigma}_{22}}}{2k}+k\left(A_{-2}^{\bs{\Sigma}_{11}}-\frac{1}{2k}\right)-\frac{\rho^2}{1+4\mu^2}}_{\det\bs{\Sigma}_\infty}
+\frac{1+8\mu_{\mathcal{F}}+4\mu_{\mathcal{F}}^2}{9\mu_{\mathcal{F}}+40\mu_{\mathcal{F}}^3+16\mu_{\mathcal{F}}^5} \frac{\rho^2}{H^2} z+\order{\rho^4, z^2}\, .
\eea 
The asymptotic value $\det\bs{\Sigma}_\infty$ cannot be determined without numerical matching, since it depends on the two $\rho^2$-suppressed constants $A_{2}^{\bs{\Sigma}_{22}}$ and $A_{-2}^{\bs{\Sigma}_{11}}-1/(2k)$. The rate at which purity grows is however fully determined by the above relation, and recalling that $\gamma=1/(4\det\bs{\Sigma})$, one finds
\bea
\gamma&=\gamma_\infty -\frac{1+8\mu_{\mathcal{F}}+4\mu_{\mathcal{F}}^2}{\frac{9}{4}\mu_{\mathcal{F}}+10\mu_{\mathcal{F}}^3+4\mu_{\mathcal{F}}^5} \frac{\rho^2}{H^2} z\\
&\simeq  \gamma_\infty -\frac{\rho^2}{H^2} \frac{1}{\mu_{\mathcal{F}}^3 }z
\simeq \gamma_\infty - \frac{\rho^2}{m^2}\frac{H}{m} z
\eea  
where in the second line we have taken the limit $m\gg H$. This coincides with Eq.~(2) in the main text.

\end{document}